\documentclass[useAMS,usenatbib,usegraphicx]{mn2e}

\title[Revisiting the RRM model for $\sigma$ Ori E I.]{Revisiting the Rigidly Rotating Magnetosphere model for $\sigma$ Ori E.  I. Observations and Data Analysis\thanks{Based on observations obtained using the Narval spectropolarimeter at the Observatoire du Pic du Midi (France), which is operated by the Institut National des Sciences de l'Univers (INSU), observations obtained at the Canada-France-Hawaii Telescope (CFHT) which is operated by the National Research Council of Canada, the Institut National des Sciences de l'Univers of the Centre National de la Recherche Scientifique of France, and the University of Hawaii, and observations collected at the European Organisation for Astronomical Research
in the Southern Hemisphere, Chile, under program ID 62.H-0319 and during the
FEROS Commissioning run II. } }
\author[M.E. Oksala et al.]{M.E. Oksala$^{1,2}$ \thanks{E-mail:meo@udel.edu}, G.A. Wade$^2$, R.H.D. Townsend$^{3}$, S.P. Owocki$^{1}$, O. Kochukhov$^{4}$,
\newauthor
C. Neiner$^{5}$, E. Alecian$^{5}$, J. Grunhut$^{2,6}$, and the MiMeS Collaboration \\
$^{1}$Bartol Research Institute, Department of Physics and Astronomy, University of Delaware, Newark, DE 19716, USA\\
$^{2}$Department of Physics, Royal Military College of Canada, P.O. Box 17000, Station Forces, Kingston, Ontario, Canada\\
$^{3}$Department of Astronomy, University of Wisconsin-Madison, 2535 Sterling Hall, 475 N Charter Street, Madison, WI 53706, USA \\
$^{4}$Department of Physics and Astronomy, Uppsala University, Box 516, Uppsala 75120, Sweden \\
$^{5}$LESIA, UMR 8109 du CNRS, Observatoire de Paris, UPMC, Universit\'e Paris Diderot, 5 place Jules Janssen, 92195, Meudon Cedex, France \\
$^{6}$Department of Physics, Engineering Physics \& Astronomy, Queen's University, Kingston, Ontario, Canada\\\
}

\begin{document}

\date{\today}

\pagerange{\pageref{firstpage}--\pageref{lastpage}} \pubyear{2011}

\maketitle

\label{firstpage}

\begin{abstract}

We have obtained 18 new high-resolution spectropolarimetric observations of the B2Vp star $\sigma$ Ori E with both the Narval and ESPaDOnS spectropolarimeters.  The aim of these observations is to test, with modern data, the assumptions of the Rigidly Rotating Magnetosphere (RRM) model of Townsend \& Owocki (2005), applied to the specific case of $\sigma$ Ori E by Townsend et al. (2005).  This model includes a substantially offset dipole magnetic field configuration, and approximately reproduces previous observational variations in longitudinal field strength, photometric brightness, and H$\alpha$ emission.  We analyze new spectroscopy, including H~{\sc i}, He~{\sc i}, C~{\sc ii}, Si~{\sc iii} and Fe~{\sc iii} lines, confirming the diversity of variability in photospheric lines, as well as the double S-wave variation of circumstellar hydrogen.  Using the multiline analysis method of Least-Squares Deconvolution (LSD), new, more precise longitudinal magnetic field measurements reveal a substantial variance between the shapes of the observed and RRM model time-varying field.  The phase resolved Stokes~$V$ profiles of He{\sc i} 5876 \AA~and 6678 \AA~lines are fit poorly by synthetic profiles computed from the magnetic topology assumed by Townsend et al. (2005). These results challenge the offset dipole field configuration assumed in the application of the RRM model to $\sigma$ Ori E, and indicate that future models of its magnetic field should also include complex, higher-order components.
\end{abstract}


\section{Introduction}

Hot, massive stars are not expected to to harbor magnetic fields due to the absence of strong convection zones in their outer envelopes.  However, peculiar A and B (Ap/Bp) stars are known to possess strong, organized magnetic fields with strengths up to tens of kG.  The hottest of these stars, the He-strong stars, are main sequence B2 stars with enhanced and often variable helium abundance.

The helium-strong B2Vp star $\sigma$ Orionis E (HD 37479) is one of the most well-studied variable massive stars.  This bright star is characterized by rapid rotation, a strong magnetic field, and variable circumstellar H$\alpha$ emission.  Observations show modulation according to the 1.19 day rotation period in longitudinal magnetic field (Landstreet \& Borra 1978), H$\alpha$ emission (Walborn 1974), He line strength (Pedersen \& Thomsen 1974), photometry (Hesser et al. 1976), UV line strength (Smith \& Groote 2001), 6 cm radio emission (Leone \& Umana 1993), and linear polarization (Kemp \& Herman 1977).  $\sigma$ Ori E has long been known as \textit{the} prototypical magnetic Bp star.

Over three decades ago, Landstreet \& Borra (1978) discovered the strong magnetic field of $\sigma$ Ori E, the first detection in a helium-strong star.  Using the circular polarization in the wings of the H$\beta$ line, the eight longitudinal magnetic field (B$_{\ell}$) measurements, when phased with the photometric ephemeris of Hesser et al. (1976), form a curve that appears nearly sinusoidal, varying from -2.2 kG to 2.8 kG.  Assuming a centered dipole magnetic field, the authors estimated a stellar surface field strength at the magnetic poles of $\sim$10 kG.  As a follow-up study to this seminal paper, Bohlender et al. (1987) observed ten helium-strong stars, presenting eight H$\beta$ and four He~{\sc i} 5876 \AA~circular polarization measurements of $\sigma$ Ori E.  The additional data confirms the periodicity and field amplitude of Landstreet \& Borra (1978).  

Walborn \& Hesser (1976) suggested that this star was in a mass-transfer binary system with an unseen companion.  However, the authors also discussed $\sigma$ Ori E as a possible application of the single-star oblique rotator model proposed by Stibbs (1950).  With the landmark discovery of the star's field, Landstreet \& Borra (1978) proposed a framework which depicts $\sigma$ Ori E as a magnetic oblique rotator with plasma trapped in a magnetosphere.  Theoretical efforts to establish a viable model to describe the physical phenomena included work by Nakajima (1985), Shore \& Brown (1990), Bolton (1994), Short \& Bolton (1994), Preuss et al. (2004), and most recently, Townsend \& Owocki (2005).  

The Rigidly Rotating Magnetosphere (RRM) model was developed by Townsend \& Owocki (2005) to analytically describe the circumstellar plasma structure of a rapidly rotating star in which the magnetic field overpowers the stellar wind, causing plasma to become trapped in a magnetosphere that co-rotates with the star.  The strong magnetic field spins up and confines the wind plasma, keeping it rigidly rotating well beyond the Kepler co-rotation radius, but also held down against the net outward gravito-centrifugal force.  The geometry of the magnetosphere depends on the obliquity angle $\beta$ and the angular velocity $\Omega$ of the star (Townsend 2008).  The observed variability also depends on the star's inclination angle, $i$.  

Townsend et al. (2005) applied the RRM model to the specific case of $\sigma$ Ori E, adopting the parameters $i = 75\degr$, $\Omega = 0.5\Omega_{c}$, and $\beta = 55\degr$.  $\Omega_{c}$ is the velocity at which the surface centrifugal force at the equator balances with gravity.  The benefit of choosing such a well-studied star is the ability to robustly confront the model with a variety of observations.  Model parameter determinations were strongly driven by observational constraints.  Motivated by the unequal primary and secondary minima in the observed photometric light curve, as well as the unequal strength of the blue and red-shifted H$\alpha$ emission, the authors defined an offset vector, $\bmath{a}$, to describe the displacement of the dipole magnetic center from the stellar center.  In units of the polar radius R$_{\rm{p}}$, the dipole magnetic field was significantly offset to $\bmath{a}=(-0.041, 0.30, 0.029)$ (corresponding to a 0.3 R$_{\rm{p}}$ displacement perpendicular to the rotating and magnetic axes and a $-$0.5 R$_{\rm{p}}$ displacement along the magnetic axis).  In confirmation of original suggestions by Landstreet \& Borra (1978), the RRM model adopted for $\sigma$ Ori E produces two regions of enhanced plasma density (clouds), located at the intersections of the magnetic and rotational equators.  A substantial success of the model was its ability to qualitatively reproduce the observed H$\alpha$ variations of the FEROS spectra previously analyzed by Reiners et al. (2000).  The model photometric light curve agrees with the observed eclipse depths and relative timing of the Hesser et al. (1977) Str\"{o}mgren u-band data.  The measurements of Landstreet \& Borra (1978) and Bohlender et al. (1987) correspond reasonably well to the computed model longitudinal field curve.  However, differences do exist between the model and observed data, such as the quantitative properties of the H$\alpha$ variations.  The excess photometric brightness at rotational phase $\sim$0.6 is likely a result of photospheric features not considered by this purely circumstellar model, as shown by Krti\u{c}ka et al. (2007) for HD~37776.

The goal of this project is a re-evaluation of the RRM model using modern data to test the original assumptions of Townsend et al. (2005), and to resolve the remaining discrepancies between the data and the model.  In this first paper, we present new spectropolarimetric observations of $\sigma$ Ori E.  Section 2 describes the observations and data reduction.  Section 3 explains the magnetometry analysis and results.  Section 4 explores the spectroscopic properties.  We discuss in Section 5 the RRM model's ability to accurately describe $\sigma$ Ori E in light of these new observations.  Section 6 presents a summary of this paper, as well as future work.

\section{Observations}

We obtained a total of 18 high-resolution (R=65000) broadband (370-1040~nm) spectra of $\sigma$ Ori E.  Sixteen spectra were obtained in November 2007 with the Narval spectropolarimeter on the 2.2m Bernard Lyot telescope (TBL) at the Pic du Midi Observatory in France.  The remaining two spectra were obtained in February 2009 with the spectropolarimeter ESPaDOnS on the 3.6-m Canada-France-Hawaii Telescope (CFHT), as part of the Magnetism in Massive Stars (MiMeS) Large Program (Wade et al. 2010).   The observation log is reported in Table \ref{magfield}.  Reduction was performed at both observatories with the Libre-ESpRIT package (Donati et al. 1997), which yields the Stokes~$I$ (intensity) and Stokes~$V$ (circular polarization) spectrum, as well as the null spectrum (N), which diagnoses any spurious contribution to the polarization measurement.  Four consecutive sub-exposures are combined using double ratios to produce one polarization spectrum.  Each resultant Stokes~$I$ spectrum was normalized by individual order using a polynomial fit to continuum points.  Each Stokes~$V$ spectrum was normalized by the same polynomial continuum fit, producing a $V/I_{c}$ spectrum.

\begin{table*}
\centering
\caption{Log of spectropolarimetric observations and magnetic field measurements.  All but the last two rows of data were taken with the Narval spectrpolarimeter.  The remaining two observations were taken with the ESPaDOnS spectropolarimeter.  Column 1 gives the heliocentric Julian date of mid-observation.  Column 2 the rotational phase according to the adopted ephemeris of Townsend et al. (2010).  Column 3 the total exposure time of the observation. Column 4 the peak signal to noise ratio per 1.8 km~s$^{-1}$ per spectral pixel in the reduced spectrum.  Column 5 the signal to noise ratio in the LSD profile. Columns 6, 7 and 8 the longitudinal magnetic field computed from individual line Stokes~$V$ profiles for H$\beta$, He {\sc i} 6678 \AA, and He {\sc i} 5876 \AA, along with their associated formal errors.  Columns 9 and 10 the longitudinal magnetic field computed from LSD mean Stokes~$V$ and N profiles, along with their associated formal errors.  }
\begin{tabular}{lcccrrrrrr}
\hline \hline
 & & & &  & H$\beta$~~~~~~ & He~{\sc i} 6678 \AA & He~{\sc i} 5876 \AA & ~~~~LSD~V~~~ & ~~~~~LSD~N~~    \\
HJD & Phase & Exp. & Peak & LSD & $B_\ell$ $\pm$ $\sigma_B$~~~& $B_\ell$ $\pm$ $\sigma_B$~~ & $B_\ell$ $\pm$ $\sigma_B$~~~& $B_\ell$ $\pm$ $\sigma_B~~$ & $B_\ell$ $\pm$ $\sigma_B$~ \\ 
(2450000+)  & & Time (s)   & S/N  &S/N& ~~~~~(G)~~~~~ & ~~~~~(G)~~~~~ & ~~~~~(G)~~~~~ & ~~~~~(G)~~~~~ & ~~~~~(G)~~~~ \\ \hline
4416.48759 & 0.672 & 3600 & 583  &  5351  &  +1933 $\pm$ 198       & +1676 $\pm$ 170   & +2107 $\pm$ 186     &  +1986 $\pm$ 78       & $-$13 $\pm$ 78 \\
4417.48739 & 0.511 & 3600 & 605  &  5736  &  +696 $\pm$ 189       & +746 $\pm$ 178      & +1137 $\pm$ 199    &  +1012 $\pm$ 76       & +89 $\pm$ 76 \\
4418.48758 & 0.351 & 3200 & 443  &  4101  &  $-$647 $\pm$ 229       & $+$26 $\pm$ 224     & $-$271 $\pm$ 247     &  $-$286 $\pm$ 102    & $-$23 $\pm$ 102 \\
4421.63402 & 0.993 & 4800 & 558  &  4165  &  +270 $\pm$ 164        & +716 $\pm$ 153    & +435 $\pm$ 147     &  +766 $\pm$ 75          & +244 $\pm$ 74 \\
4422.53892 & 0.753 & 4800 & 725  &  5402  &  +2306 $\pm$ 154       & +2022 $\pm$ 126     & +2151 $\pm$ 120     &  +2350 $\pm$ 58        & $-$38 $\pm$ 58\\
4426.54364 & 0.116 & 4800 & 568  &  7607  &  $-$1371 $\pm$ 185     & $-$908 $\pm$ 174   & $-$1602 $\pm$ 169     &  $-$1395 $\pm$ 80     & +206 $\pm$ 80\\
4429.47771 & 0.580 & 4800 & 558  &  6884  &  +1680 $\pm$ 191       & +2071 $\pm$ 188     & +1439 $\pm$ 189    &  +1430 $\pm$ 82        & $-$120 $\pm$ 81\\
4429.53649 & 0.629 & 4800 & 287  &  5512  &  +1927 $\pm$ 220       & +1959 $\pm$ 204      & +1283 $\pm$ 204     &  +2021 $\pm$ 164      & $-$194 $\pm$ 163 \\
4429.61850 & 0.698 & 4800 & 513  &  5385  &  +1786 $\pm$ 205       & +2087 $\pm$ 174     & +1722 $\pm$ 173     &  +2078 $\pm$ 85        & $-$88 $\pm$ 84\\
4432.55955 & 0.168 & 4800 & 582  &  2570  &  $-$1847 $\pm$ 175      & $-$1925 $\pm$ 171      & $-$1537 $\pm$ 168     &  $-$1878 $\pm$ 79    & +19 $\pm$ 79\\
4432.62834 & 0.226 & 4800 & 500  &  4891  &  $-$1492 $\pm$ 193     & $-$1190 $\pm$ 195      & $-$1465 $\pm$ 194     &  $-$1650 $\pm$ 90    & $-$139 $\pm$ 90\\
4432.68693 & 0.275 & 4800 & 520  &  5573  &  $-$1141 $\pm$ 187      & $-$674 $\pm$ 182      & $-$999 $\pm$ 186    &  $-$1179 $\pm$ 86    & $-$83 $\pm$ 86\\
4433.49259 & 0.951 & 4640 & 721  &  4866  &  +874 $\pm$ 133          & +1235 $\pm$ 116    & +1116 $\pm$ 112    &  +1518 $\pm$ 57        & +94 $\pm$ 57\\
4433.54897 & 0.999 & 4640 & 737  &  5045  &  $-$194 $\pm$ 128        & +470 $\pm$ 119    & +503 $\pm$ 114     &  +577 $\pm$ 58           & +121 $\pm$ 57\\
4433.60532 & 0.046 & 4640 & 703  &  7066  &  $-$832 $\pm$ 131       & $-$562 $\pm$ 126     & $-$534 $\pm$ 121     &  $-$605 $\pm$ 60       & +39 $\pm$ 60\\
4433.66177 & 0.093 & 4640 & 719  &  7006  &  $-$1612 $\pm$ 144     & $-$1103 $\pm$ 132   & $-$1234 $\pm$ 131     &  $-$1113 $\pm$ 62     & +31 $\pm$ 62\\
4875.83320 & 0.400 & 1600 & 416  &  6983  &  +446 $\pm$ 221          & +970 $\pm$ 213   & +275 $\pm$ 231    &  +507 $\pm$ 98           & $-$46 $\pm$ 97\\
4878.71470 & 0.820 & 1600 & 748  &  7045  &  +2080 $\pm$ 164        & +1943 $\pm$ 114    & +1735 $\pm$ 116     &  +2345 $\pm$ 54         &   +5 $\pm$ 53\\
\hline \hline
\end{tabular}
\label{magfield}
\end{table*}

Additionally, 29 FEROS echelle spectra were used in the spectroscopic analysis of $\sigma$ Ori E (Kaufer et al. 1999; See Section 4).   These data were taken in November and December 1998, partially during the commissioning of the FEROS spectrograph on the ESO 1.52-m telescope at La Silla.  These observations are further described by Reiners et al. (2000).

\section{Magnetic Field Measurements}

\subsection{Least Squares Deconvolution}

We used the multiline analysis method of Least-Squares Deconvolution (LSD; Donati et al. 1997) to produce mean Stokes~$I$ and~$V$ profiles from our spectra for calculation of high-precision longitudinal magnetic field measurements.  Silvester et al. (2009) describe the details of the LSD method as applied to B stars.  To determine appropriate parameters for the LSD mask, we studied the effect of various combinations of $T_{\rm{eff}}$, $\log g$, line depth cutoff level, spectral ranges, and chemical elements on the LSD results.   

Starting with Vienna Atomic Line Database (VALD; Kupka et al. 1999) ``extract stellar'' line lists, we produced synthetic LTE spectra using the spectral synthesis code Synth3 described by Kochukhov (2007).  For consistency, each synthetic spectrum was compared to the same observation, taken at phase 0.672 during minimum helium absorption.  The spectra used initial parameters of $T_{\rm{eff}} = 23000$~K, $\log g = 4.0$, as suggested by Cidale et al. (2007), although spectra were also computed by varying both temperature (by $\pm$ 3000K) and $\log g$ (by $\pm$ 0.5).  The abundance parameters of the synthetic spectra were adjusted until good visual agreement was reached with the observed spectra.  Although the He lines in the red part of the spectrum are affected by Stark broadening and NLTE effects, the bluer helium lines were best fit by a helium abundance of $\log$(N$_{\rm{He}}$/N$_{\rm{Tot}}$) = $-$0.1.  Although lines of certain metals may be variable, solar abundances fit well for a majority of rotational phases.  At this stage, we set the chemical abundances and proceeded to determination of an LSD line mask.

\begin{figure}
\centering
\includegraphics[width=3.4 in]{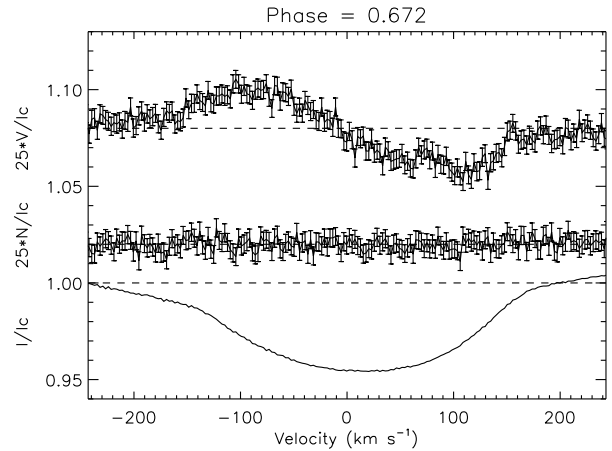}
\caption{LSD Stokes~$I$ (bottom), null $N$ (middle), and Stokes~$V$ (top) profiles of $\sigma$ Ori E at rotational phase 0.672.  The profiles are expanded by the indicated factor.  The error bars on the null and Stokes~$V$ data are propagated from the formal error bars associated with each pixel.  For visualization purposes, the null and Stokes~$V$ have been shifted upwards from 0.0.  A clear Zeeman signature is detected in the Stokes~$V$ profile, while the null profile shows no signal.}
\label{singleLSD}
\end{figure}

We next created line masks consisting of spectral line wavelengths, line depths, and atomic data for each transition contributing to the spectrum.  The normalized observational spectra were then processed using this line mask in the LSD procedure.  The output of LSD provides mean Stokes~$I$ and~$V$ profiles, mean null profiles, as well as statistics related to the procedure fit and the output profiles.  The LSD procedure propagates the formal error associated with each pixel throughout the deconvolution process.  An example of the LSD profiles for the final line mask is displayed for phase 0.672 in Fig. \ref{singleLSD}.  We adopted a velocity bin of 2.6 km~s$^{-1}$ for the calculation of profiles.    The velocity range was set from $-$243 km~s$^{-1}$ to +343 km~s$^{-1}$ to include the line itself and check for any signal that may be outside the profile.   

To test the sensitivity of the LSD line mask, we conducted an experiment to explore the impact of varying certain parameters.  A large set of masks were created and run through the LSD procedure to view any differences.  Inclusion or exclusion of certain lines or elements is an important effect on the quality of the LSD profiles.  When we excluded helium lines from our line mask, the resulting profiles were extremely noisy, as a result of the weakness of metal lines.  Including lines located in the spectral regions overlapping the higher-order Balmer series ($\lambda <$ 4000 \AA) or regions dominated by telluric lines ($\lambda >$ 8000 \AA) also affected the results negatively.  Varying T$_{\rm{eff}}$ by $\pm$ 3000~K and $\log g$ by $\pm$ 0.5 had little effect on the profiles.  A line depth limit of greater than 10\% of the continuum produced the cleanest LSD profiles, however changing the value did not significantly worsen the results.  At the conclusion of the investigation, the best overall line mask was chosen for the spectrum of $\sigma$ Ori E.

The final mask we selected begins with a line list for $T_{\rm{eff}} = 23000$~K, log $g$ = 4.0, and solar abundances except for helium.  The helium abundance was set to $\log$(N$_{\rm{He}}$/N$_{\rm{Tot}}$) = $-$0.1.  A total of 332 lines contributed to the line mask, and most helium and metal lines in the range 405-800~nm were included.  Lines blended with hydrogen Balmer lines within the included range were removed, as hydrogen lines are not included in the mask.  The wavelength range was chosen in an attempt to exclude regions with many hydrogen lines in the blue part of the spectrum and strong telluric contamination in the red part of the spectrum.  To be clear, this study is not meant as a precise stellar parameter determination.  We have not modeled in detail the spectrum and are not attempting to make any assertions regarding effective temperature or gravity of this star.  The process detailed here simply attempts to determine the overall best line mask to extract clean Stokes~$I$ and $V$ profiles from the LSD procedure.

\begin{figure*}
\centering
\includegraphics[width=6.5 in]{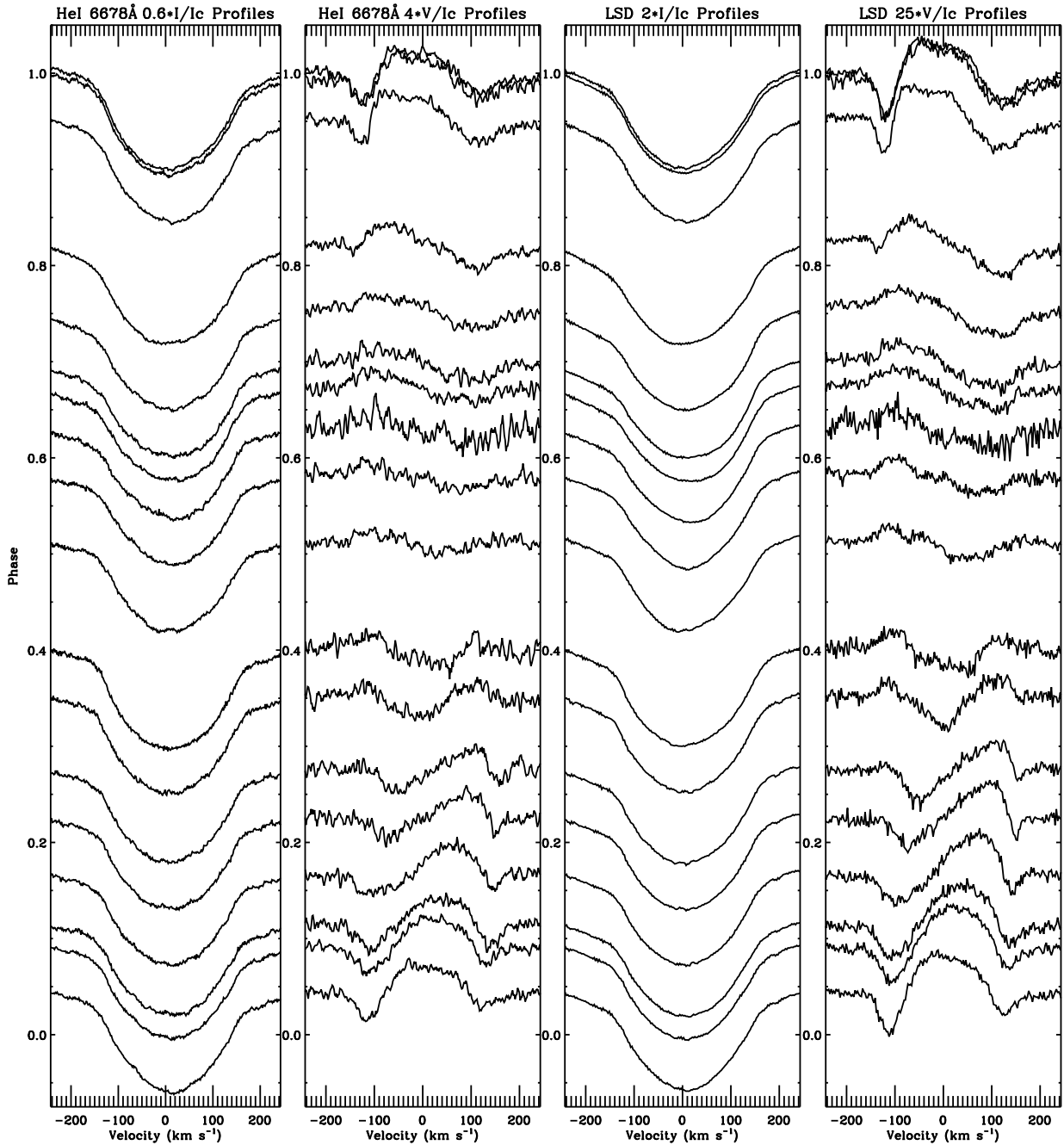}
\caption{Stokes~$I$ and Stokes~$V$  profiles of $\sigma$ Ori E for the He~{\sc i} 6678 \AA~line (left) and for the LSD fits (right).  The profiles are expanded by the indicated factor and each separate observation is shifted upwards to its proper rotational phase calculated using the ephemeris in Eq. (\ref{eqephem}).  Clear Zeeman signatures are detected in each Stokes~$V$ profile.  The He~{\sc i} 6678 \AA~profiles have been binned by 2 km~s$^{-1}$ in velocity, while the calculation of LSD profiles adopted a velocity bin of 2.6 km~s$^{-1}$. }
\label{LSD_6678}
\end{figure*}

Lack of signal in the null profiles indicates that there are no important spurious contributions to the Stokes V profiles.  Each separate LSD Stokes~$V$ profile produces a definite detection (detection probability $>$ 99.999\%) according to the criteria described by Donati et al. (1997).   Fig. \ref{LSD_6678} presents the Stokes~$I$ and~$V$ profiles for both the He~{\sc i} 6678 \AA~line, as well as the LSD mean $I$ and $V$ profiles, for each rotational phase.  The helium and LSD Stokes~$V$ profiles agree quite well, as expected as helium lines are the strongest lines contributing to the LSD mask.

\subsection{Longitudinal Magnetic Field}

\begin{figure}
\centering
\includegraphics[width=3.4 in]{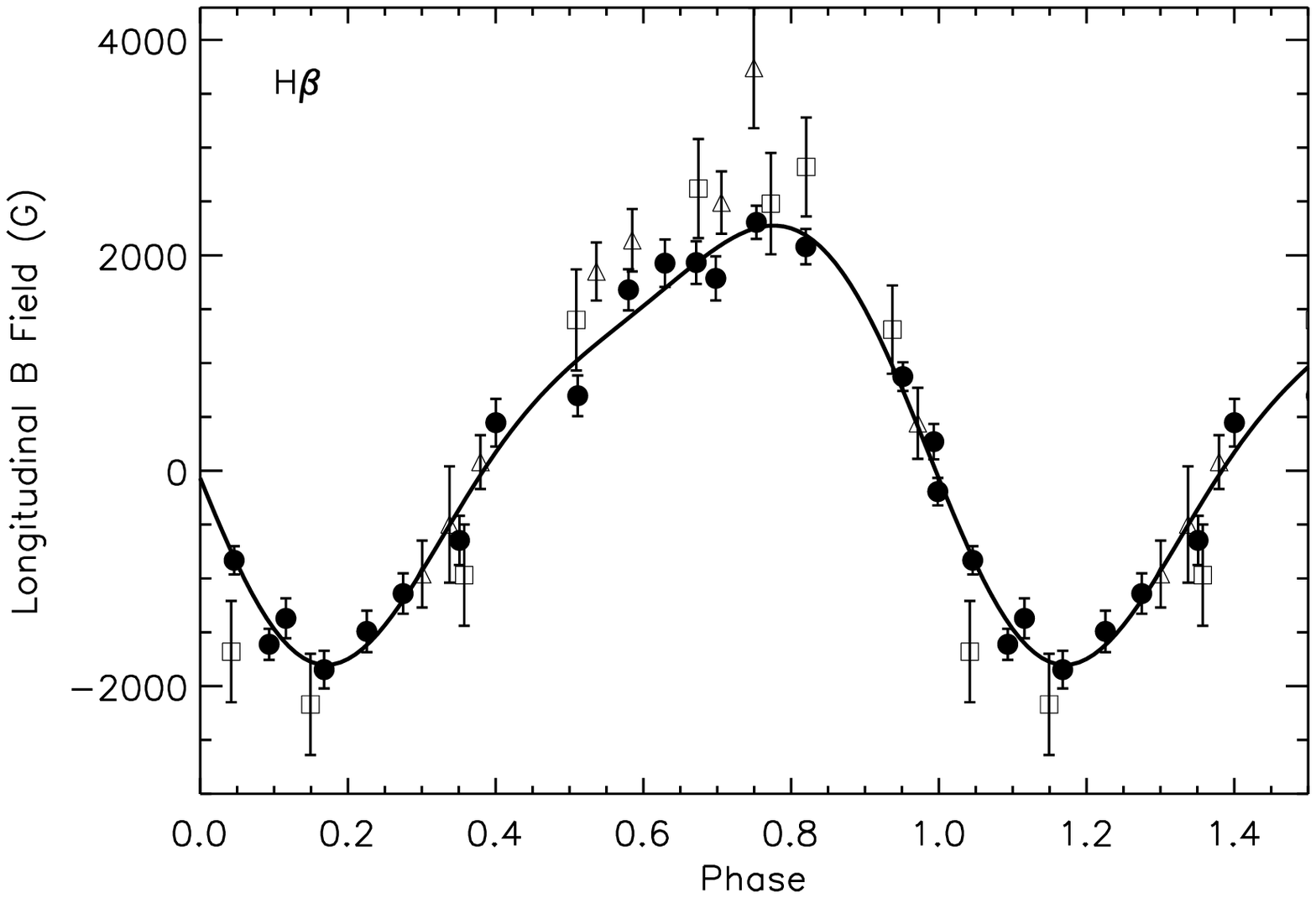}
\includegraphics[width=3.4 in]{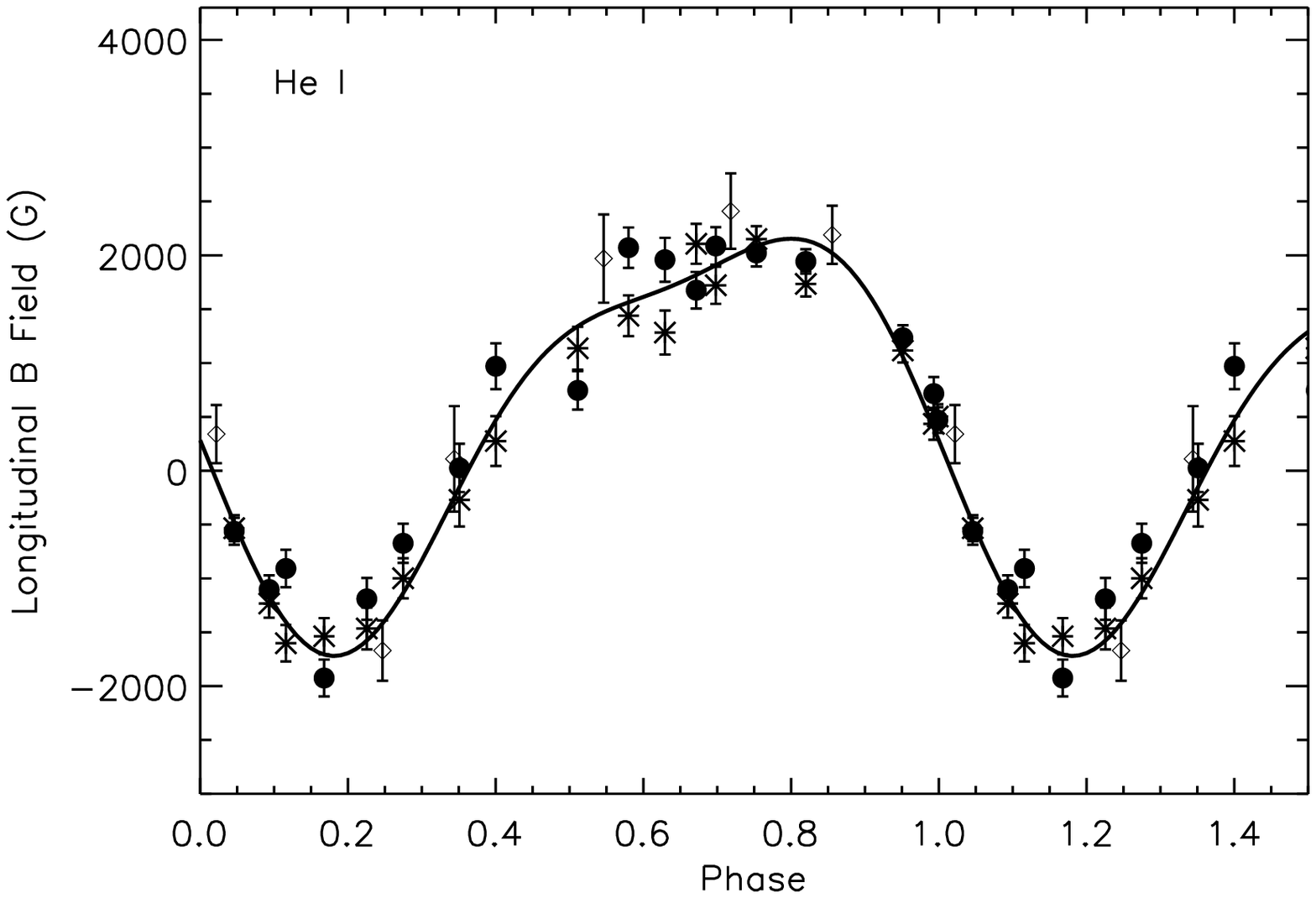}
\includegraphics[width=3.4 in]{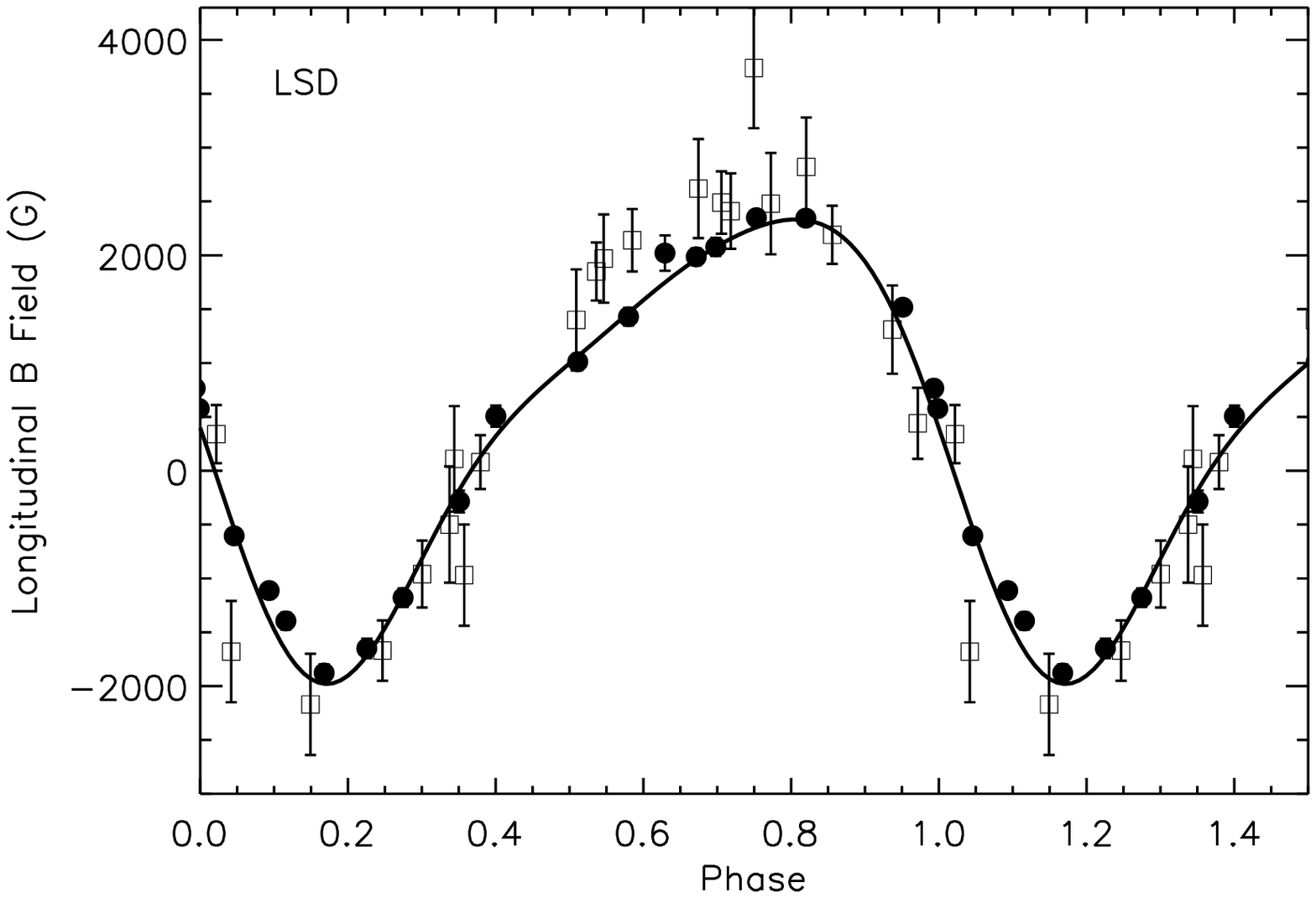}
\caption{Longitudinal magnetic field measurements for $\sigma$ Ori E.  The solid curve is the second order (H$\beta$, He~{\sc i}) or third order (LSD) least-squares sine curve fit to the CFHT and Narval spectropolarimetric data.  The data are phased according to the ephemeris of Townsend et al. (2010).  \textbf{Top:} H$\beta$ Stokes V measurements (filled circles), as well as measurements reported in Landstreet \& Borra (1978) (squares) and Bohlender et al. (1987) (triangles) for H$\beta$.  \textbf{Middle:} He~{\sc i} 6678 \AA~(filled circles) and He~{\sc i} 5876 \AA~(asterisks) Stokes V measurements, and helium line measurements from Bohlender et al. (1987) (diamonds).  \textbf{Bottom:} LSD profile measurements (filled circles), as well as the data reported in Landstreet \& Borra (1978) and Bohlender et al. (1987) (squares)  }
\label{HydBl}
\end{figure}

Historically, the longitudinal magnetic field measurements (B$_{\ell}$) of $\sigma$ Ori E were computed from the circular polarization signatures in the wings of the H$\beta$ and He~{\sc i} 5876 \AA~lines (e.g. Landstreet 1980, 1982; Borra \& Landstreet 1977).  We utilize the circular polarization and intensity spectra of H$\beta$, He~{\sc i} 6678 \AA, and He~{\sc i} 5876 \AA~to determine the B$_{\ell}$ value at each phase.  These values are reported in Table \ref{magfield} along with 1$\sigma$ uncertainties.  We calculate the longitudinal magnetic field from the first moment of Stokes~$V$: 
\begin{equation}
B_{\ell} = -2.14 \times 10^{11} \frac{ \int v V(v) dv}{\lambda g c \int [I_{c}-I(v)]dv}~~~G  ,
\label{Bleq}
\end{equation}
\noindent where $g$ is the Land\'{e} factor and $\lambda$ is the wavelength of the specified line (Donati et al. 1997; Wade et al. 2000).  The integration range (in velocity, $v$ relative to the line center-of-gravity) employed in calculating the longitudinal magnetic field was from $-$215 km~s$^{-1}$ to +215 km~s$^{-1}$ for helium lines and from $-$500 km~s$^{-1}$ to +500 km~s$^{-1}$ for H$\beta$.  These velocity ranges were chosen to be sure to include the entire Stokes~$V$ profile, but not too wide as to artificially increase the error estimates by including excess continuum.  The new measurements are plotted in Fig. \ref{HydBl}, along with the corresponding historical data from Landstreet \& Borra (1978) and Bohlender et al. (1987).  The data are phased according to the ephemeris found by Townsend et al. (2010):
\begin{equation}
JD=2442778\fd819~+~1\fd1908229E~+~1\fd44\times10^{-9}E^{2}.
\label{eqephem}
\end{equation}
\noindent This second-order ephemeris accounts for the increasing period of $\sigma$ Ori E due to rotational braking by the magnetic field.

The historical H$\beta$ measurements and new measurements from H$\beta$ Stokes V profiles agree relatively well.  There is better consistency with the He~{\sc i} lines.  The new measurements have much smaller error bars by a factor of $\sim$3, and the amplitude of the variation appears smaller than the historical data.  The amplitude difference could be a consequence of the fact that the historical measurements were taken from the wings of the line, while our measurements are mainly in the core.  A second order least-squares sine curve fit to the new H$\beta$ measurements gives a reduced $\chi^{2}$ value of 1.54.  The new helium data from He {\sc i} 5876 \AA~and 6678\AA~lines fit with a second order least-squares sine curve have a reduced $\chi^{2}$ of 2.62.  Fig \ref{HydBl} shows both the data and the corresponding fits.  When examining measurements from two different elements, it is important to remember that longitudinal magnetic field curves are affected not only by the line of sight component of the field, but also by the surface distribution of that element (Deutsch 1958; Pyper 1969).  Bohlender et al. (1987) found little difference between the hydrogen and helium measurements as far as the shape of the curve and the magnetic extrema are concerned.  We find a similar shape, but the H$\beta$ measurements best fit has extrema $\sim$ 150-250 G larger than the best fit to the helium data.

From the mean LSD profiles, we used Eq. (\ref{Bleq}) to calculate the longitudinal magnetic field, setting $g$ and $\lambda$ to the S/N-weighted mean values for all the included lines.  The integration range employed in calculating the longitudinal magnetic field was from $-$200 km~s$^{-1}$ to +200 km~s$^{-1}$, with the same considerations as for the individual line measurements.  The derived values, along with their 1$\sigma$ uncertainties, are reported in Table \ref{magfield}.  The longitudinal magnetic field is observed to change from about $-$1.9 kG to +2.4 kG.  Uncertainties for the longitudinal field values vary from 54 to 164 G, decreasing the error bars by a factor of four compared to the historical data.  The longitudinal magnetic field curve is shown in the bottom panel of Fig. \ref{HydBl}, including the historical magnetic data.  The agreement between new and historic data, phased according to the Townsend et al. (2010) ephemeris, indicates long-term stability of the large-scale magnetic field.  The photometric light curve structure shows similar stability, further indicating that the entire system, while not static, is not undergoing large-scale changes in magnetic and magnetospheric structure.  Magnetic extrema occur at phases 0.2 and 0.8, while null longitudinal fields occur at phases 0.0 and 0.4.  The shape of the magnetic field curve is quite clearly non-sinusoidal.  A third order least-squares sine curve fits the data well (reduced $\chi^{2}$ = 2.26).  The asymmetry of the longitudinal magnetic field variation may indicate a magnetic field that is not a simple oblique dipole.  

Using the Lomb-Scargle periodogram (Lomb 1976; Scargle 1982), the historical and LSD derived magnetic data gives a period of 1.190842 $\pm$ 0.000004 days, where the reported error is the 3$\sigma$ uncertainty.  Our derived period is longer than the period of 1.19081 $\pm$ 0.00001 days Hesser et al. (1977) obtained from photometric data, but consistent with a period of 1.19084 $\pm$ 0.00001 days derived from He~{\sc i} 4471 \AA~equivalent width variations in Reiners et al. (2000).  Our data span a baseline long enough that the period can be determined down to a third of a second.  However, due to the magnetic braking reported by Townsend et al. (2010), our period determination from the longitudinal magnetic field data is strictly an average period over the time frame of the observations.  In fact, when the current epoch of magnetic data are phased with the mean derived period and the HJD$_{0}$ of Hesser et al (1976), the data do not define a null longitudinal magnetic field at phase 0.0, suggested by the RRM model.  If the data are phased with the ephemeris of Townsend et al. (2010), as given in Eq. (\ref{eqephem}), the longitudinal field curve better reflects the expected phases of null fields and extrema.

\subsection{Summary}

We have derived longitudinal magnetic field measurements from Stokes~$V$ profiles of the H$\beta$, He{\sc i} 5876 \AA, and He {\sc i} 6678 \AA~lines.  These new measurements agree relatively well with historical data, although the amplitude of the variation is slightly smaller for the new data.  We also used the LSD method to produce clean, accurate mean Stokes~$I$, Stokes~$V$, and null profiles, as well as longitudinal magnetic field measurements for each of the 18 new spectropolarimetric observations of $\sigma$ Ori E.  The new measurements, when phased according to the ephemeris of Townsend et al. (2010), exhibit a non-sinusoidal periodicity (Fig. \ref{HydBl}).  The overall shape of the curve suggests a deviation from a simple oblique dipole.  Comparison with historical data indicates a long-term stability of the magnetic field structure.  In the discussion, we will compare the currently available observations with the offset dipole magnetic field configuration that Townsend et al. (2005) applied to the RRM model for $\sigma$ Ori E.

\section{Spectroscopy}

\begin{figure*}
\centering
\includegraphics[width=3.0 in]{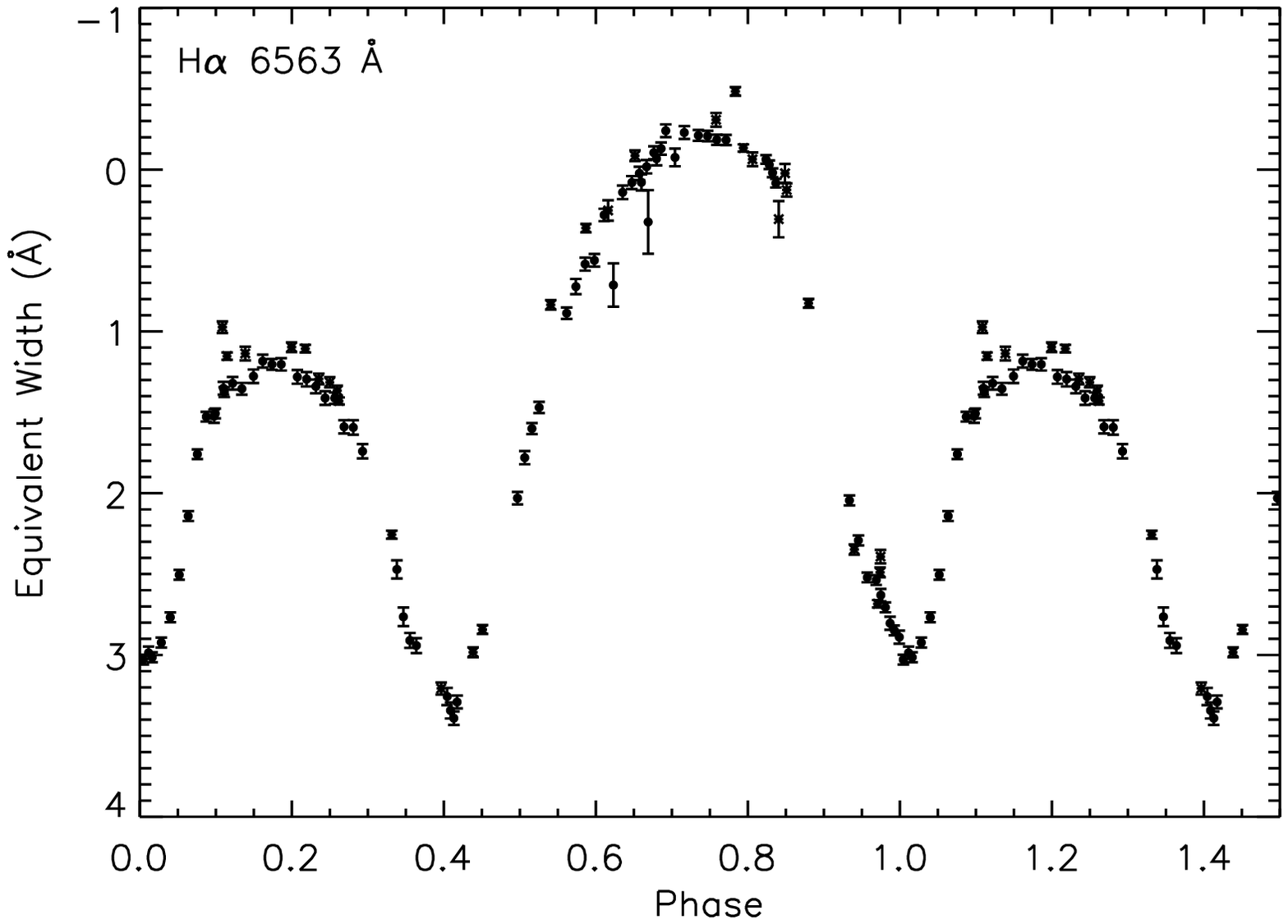}
\includegraphics[width=3.0 in]{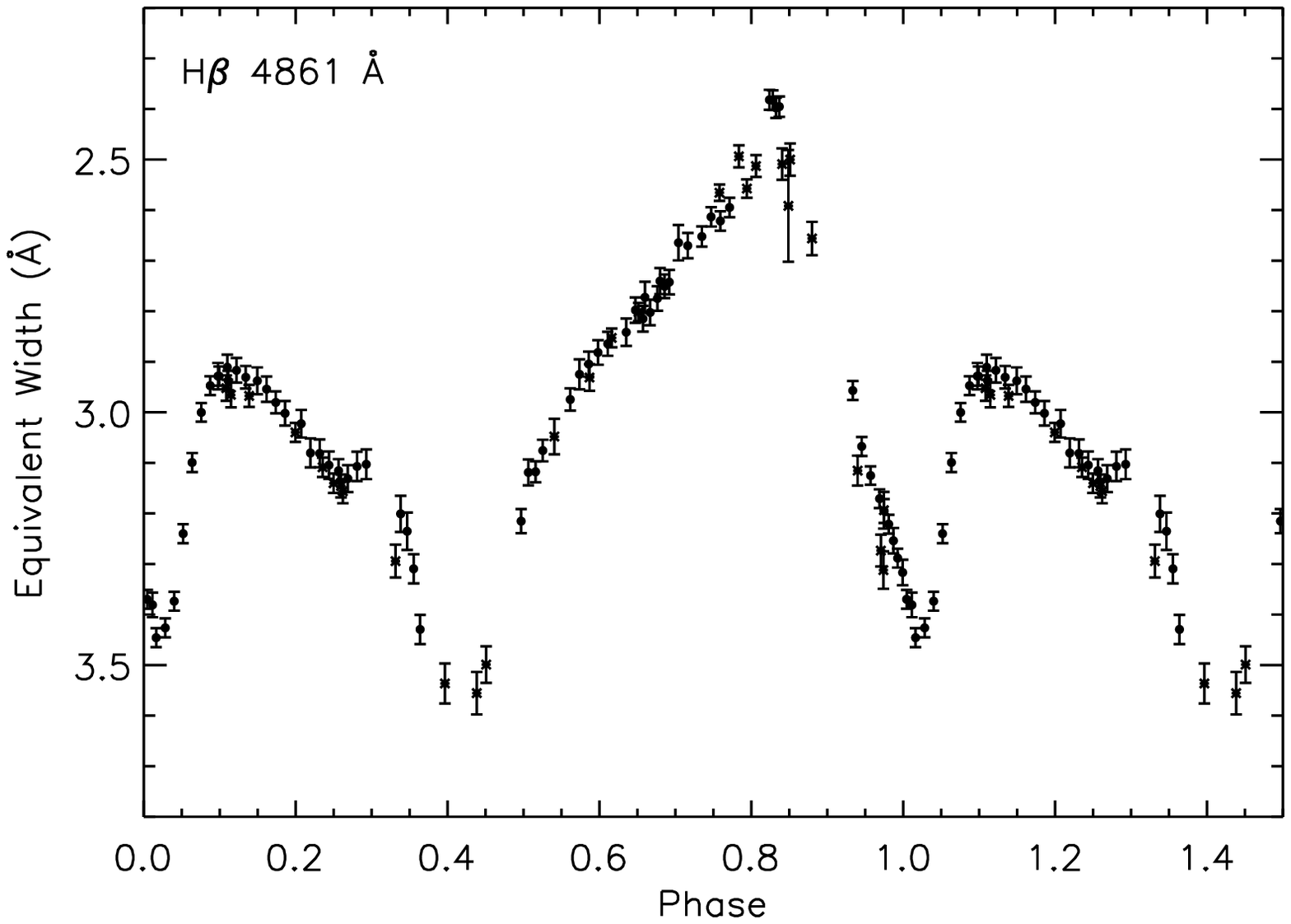}
\includegraphics[width=3.0 in]{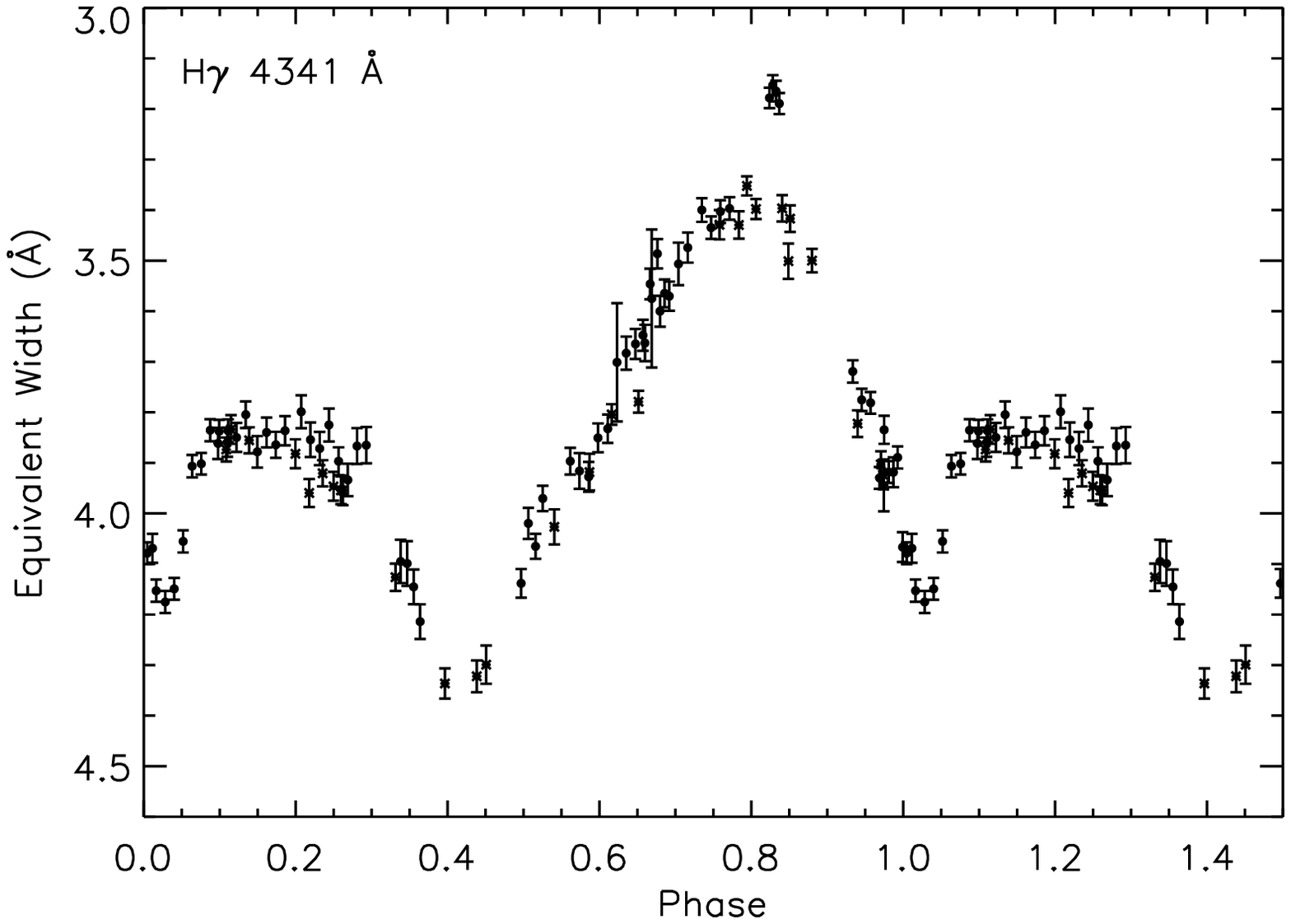}
\includegraphics[width=3.0 in]{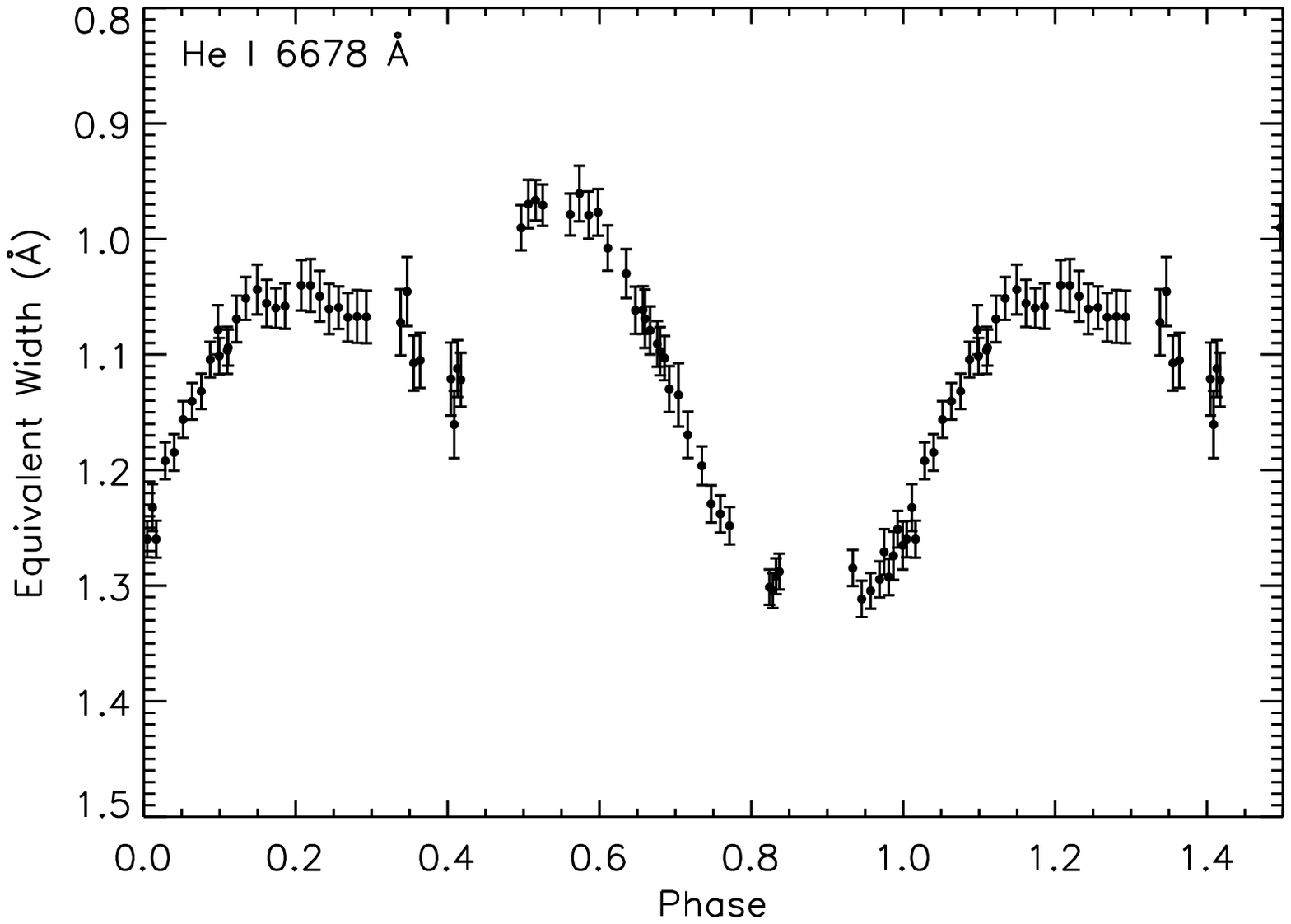}
\includegraphics[width=3.0 in]{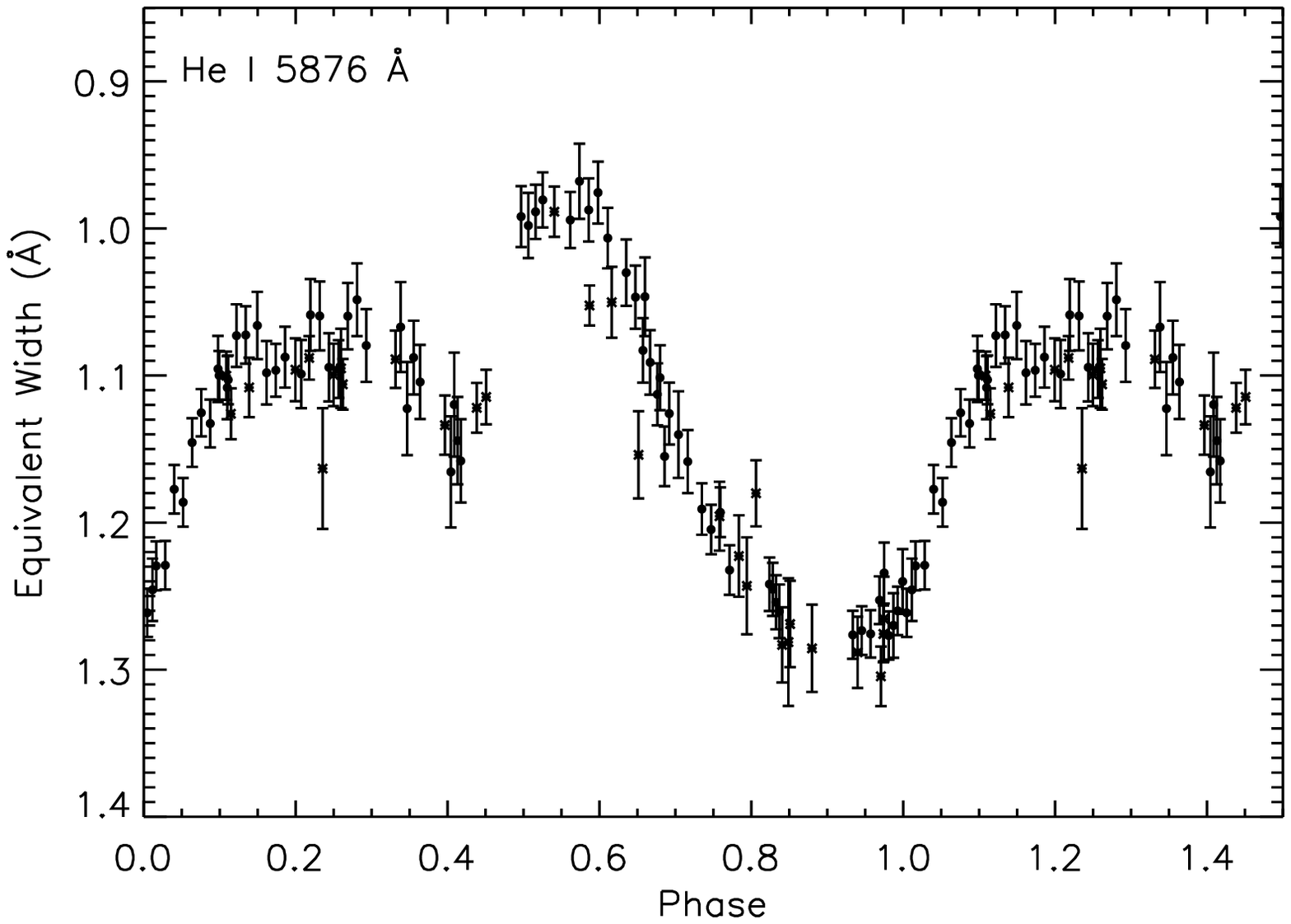}
\includegraphics[width=3.0 in]{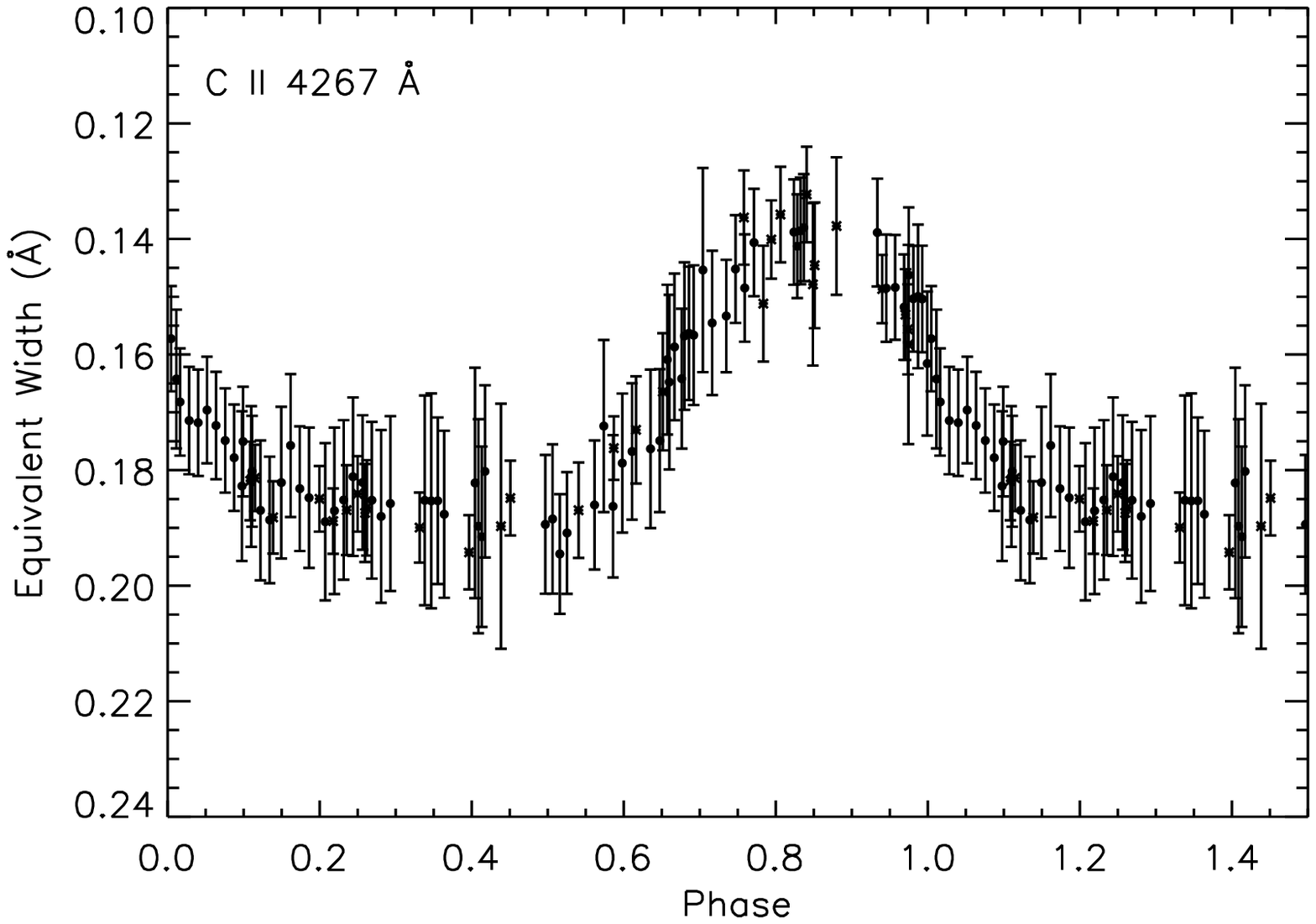}
\includegraphics[width=3.0 in]{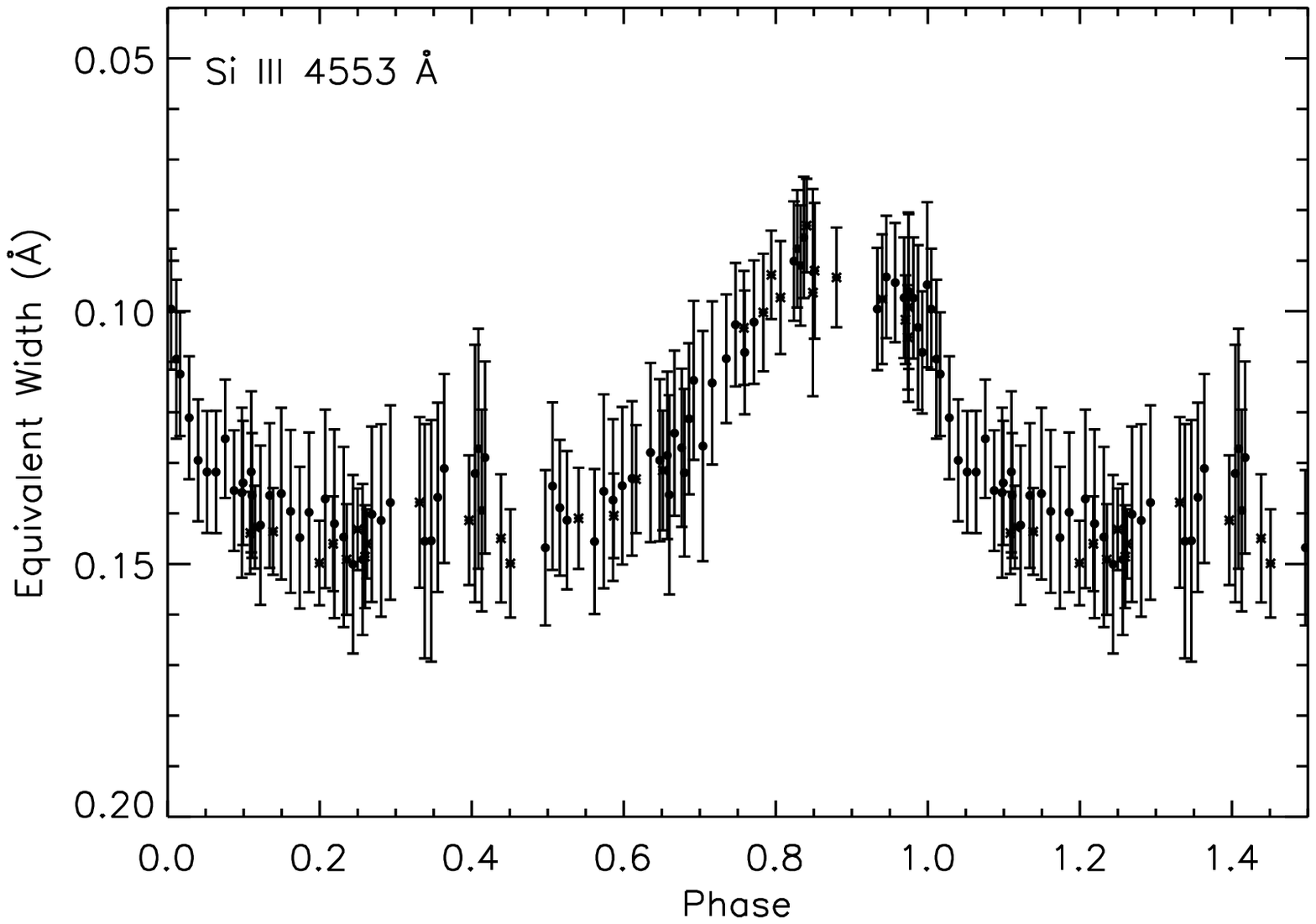}
\includegraphics[width=3.0 in]{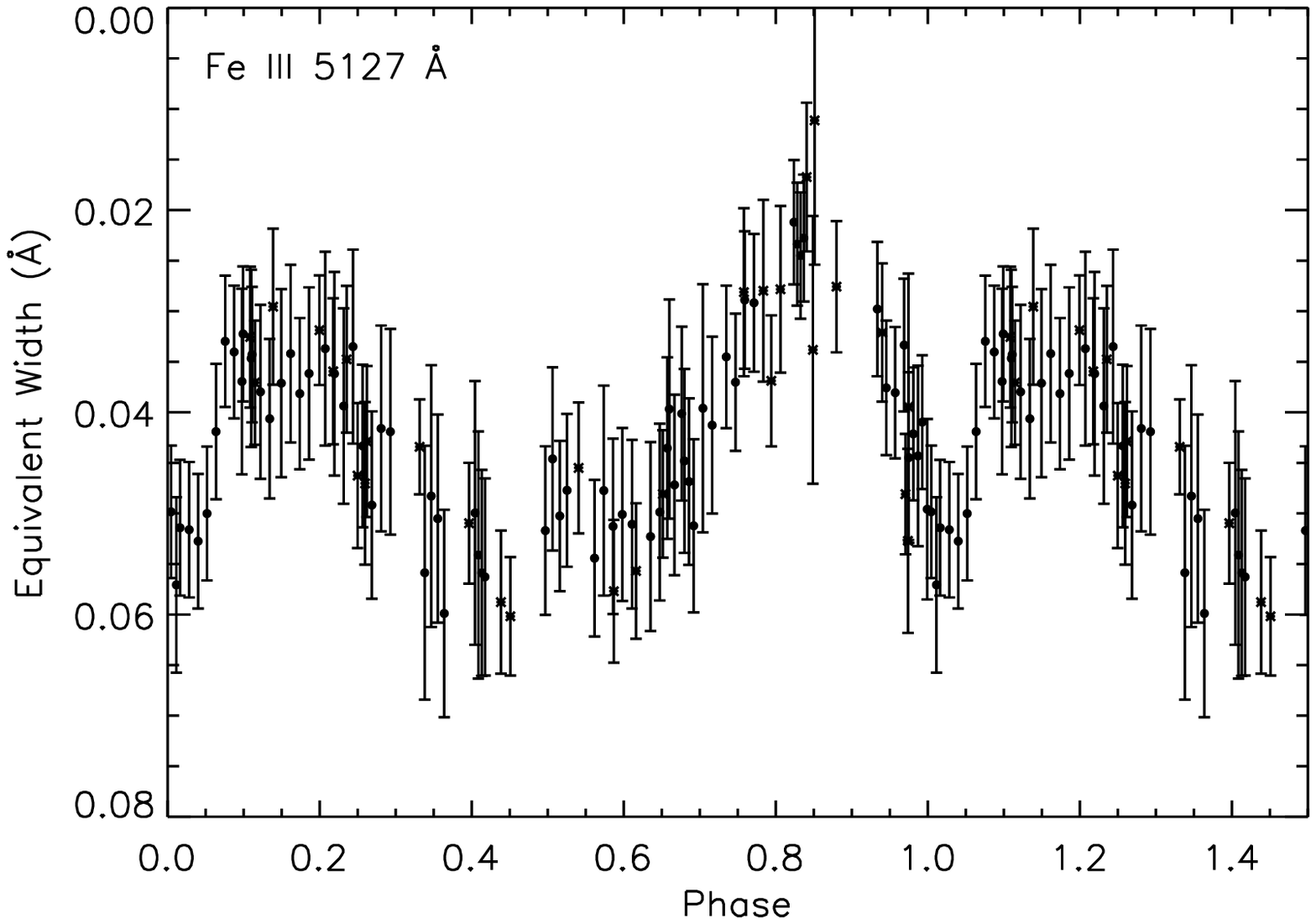}
\caption{Equivalent width curves for various spectral lines phased using the ephemeris of Townsend et al. (2010).   FEROS data was not utilized for He~{\sc i} 6678 \AA~due to a bad pixel feature in the blue wing of the line.  \textbf{Top Row}: Hydrogen H$\alpha$ (left) and H$\beta$ (right).  \textbf{Second Row}: H$\gamma$ (left) and Helium~{\sc i} 6678 \AA~(right).  \textbf{Third Row}:  Helium~{\sc i} 5876 \AA~(right) and Carbon {\sc ii} 4267 \AA~(right). \textbf{Bottom row}: Silicon~{\sc iii} 4553 \AA~(left) and Fe~{\sc iii} 5127 \AA~(right). }
\label{EW}
\end{figure*}

\begin{figure*}
\centering
\includegraphics[width=3.4 in]{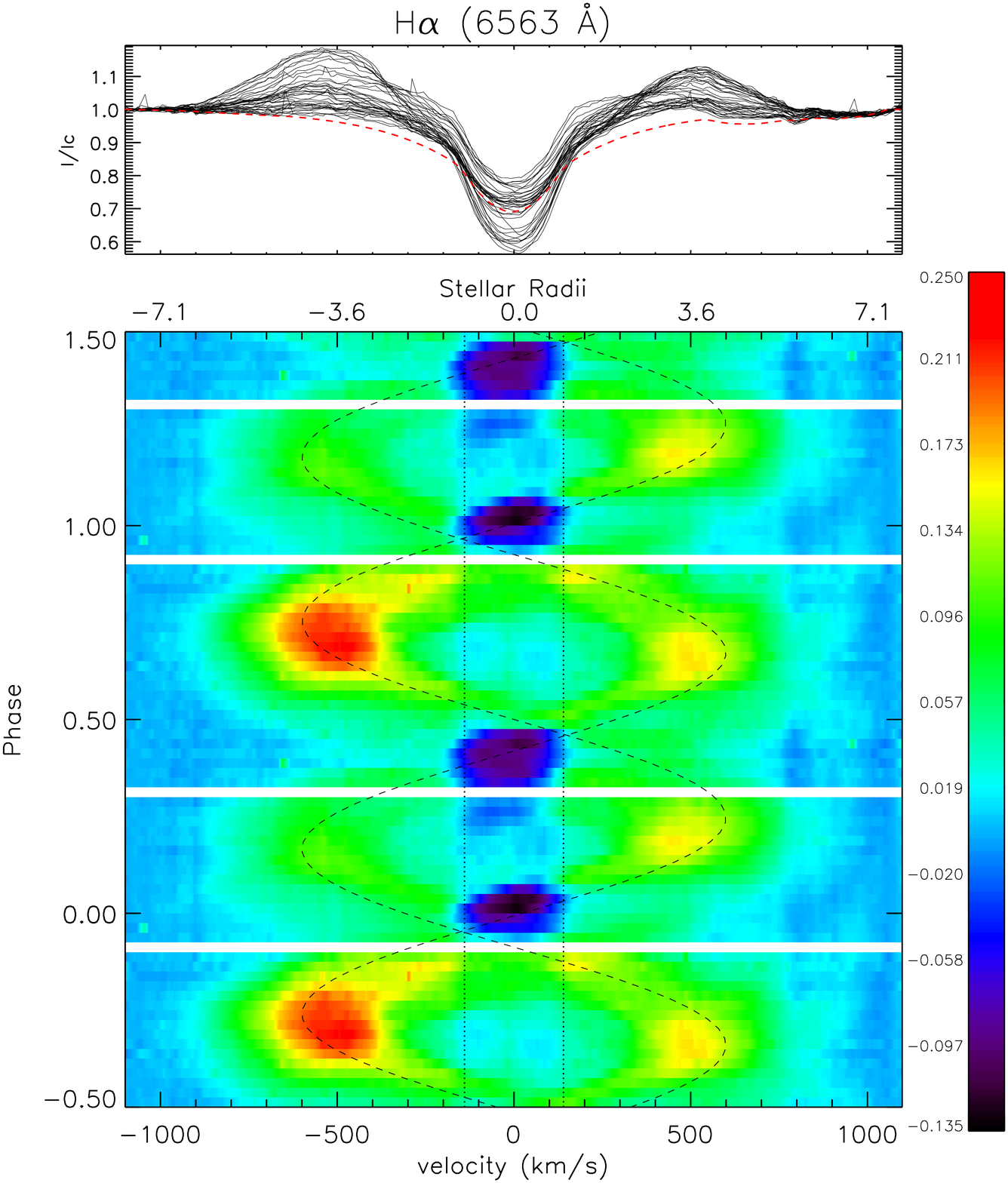}
\includegraphics[width=3.4 in]{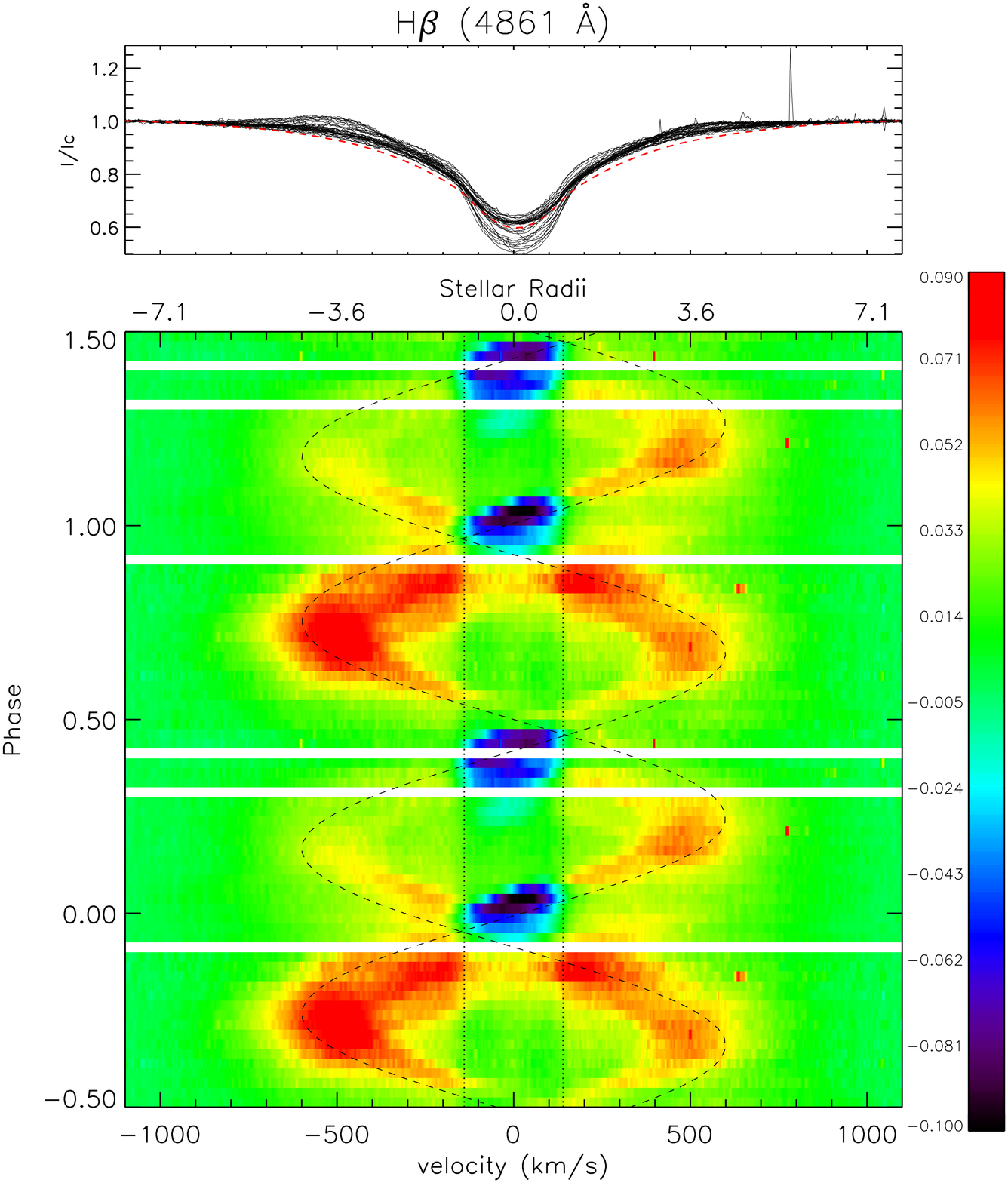}
\caption{Dynamic spectral plots for H$\alpha$ and H$\beta$, phased using the ephemeris of Townsend et al. (2010).  Spectra are plotted minus a synthetic photospheric spectrum.  Dashed vertical lines mark the stellar rotational velocity.  Dashed curves represent the motion of the emission features (clouds) with an amplitude of 600 km s$^{-1}$ and transits at rotational phases $\sim$0.0 and $\sim$0.42. }
\label{hyddynplot}
\end{figure*}

\begin{figure*}
\centering
\includegraphics[width=3.4 in]{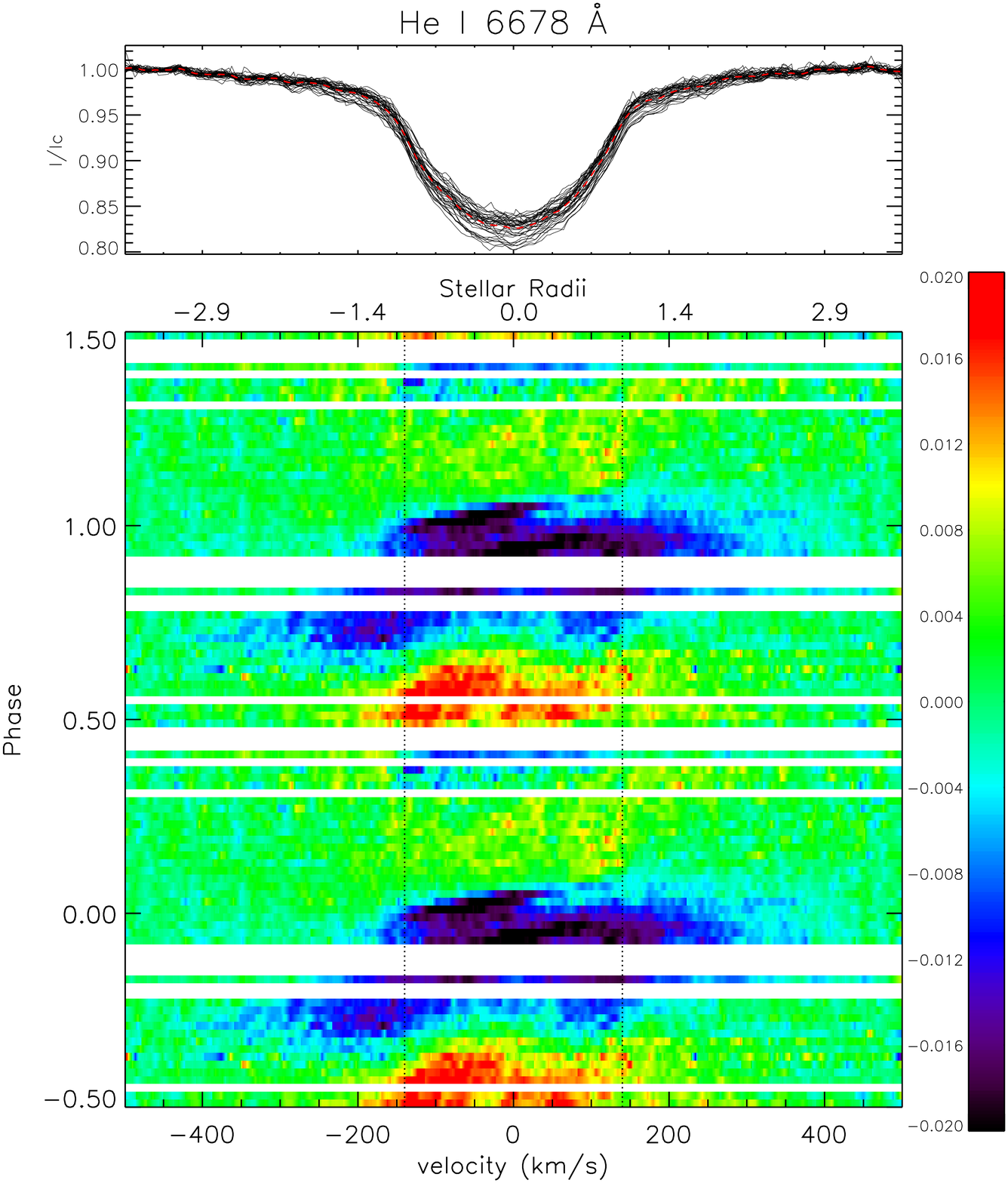}
\includegraphics[width=3.4 in]{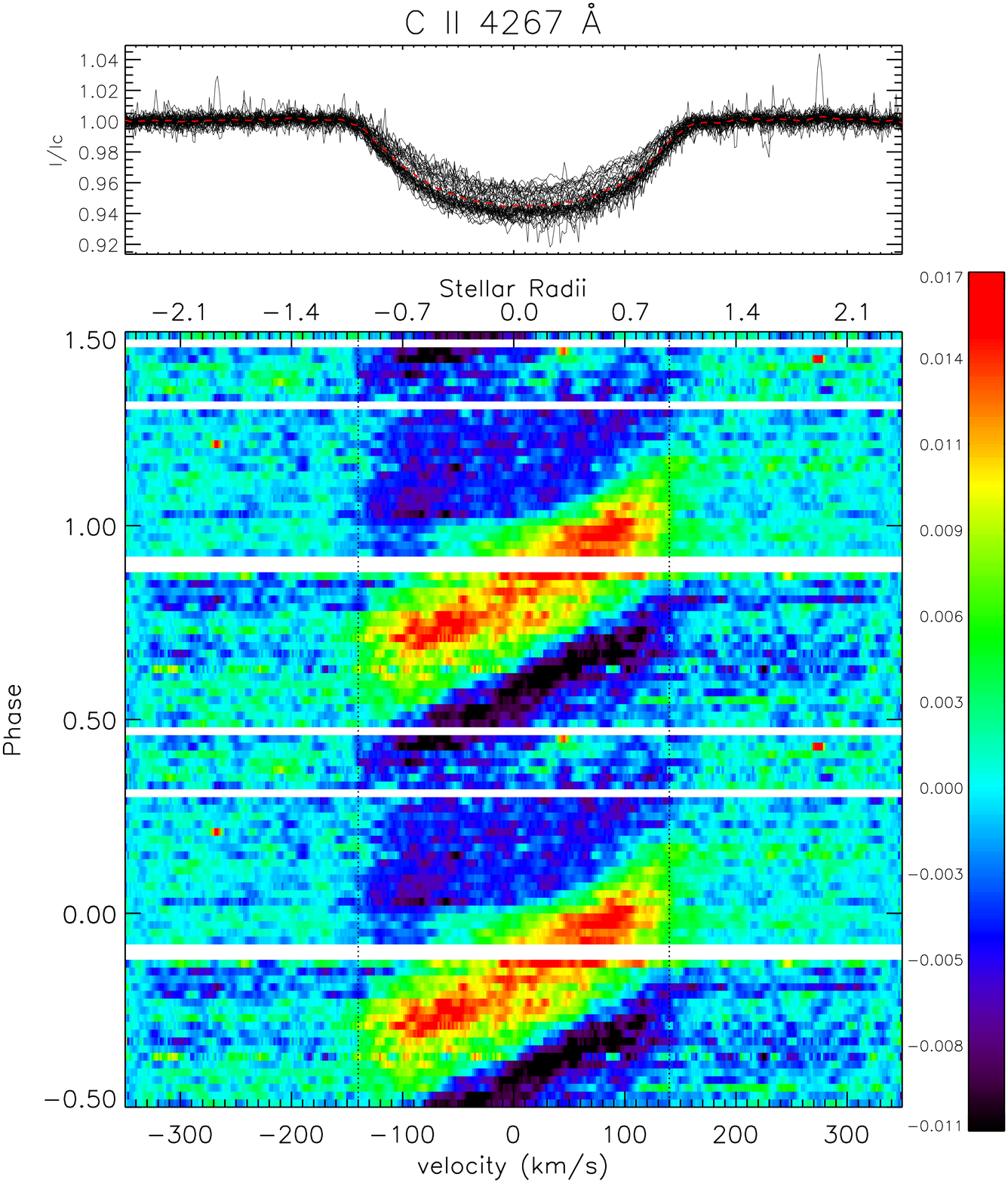}
\includegraphics[width=3.4 in]{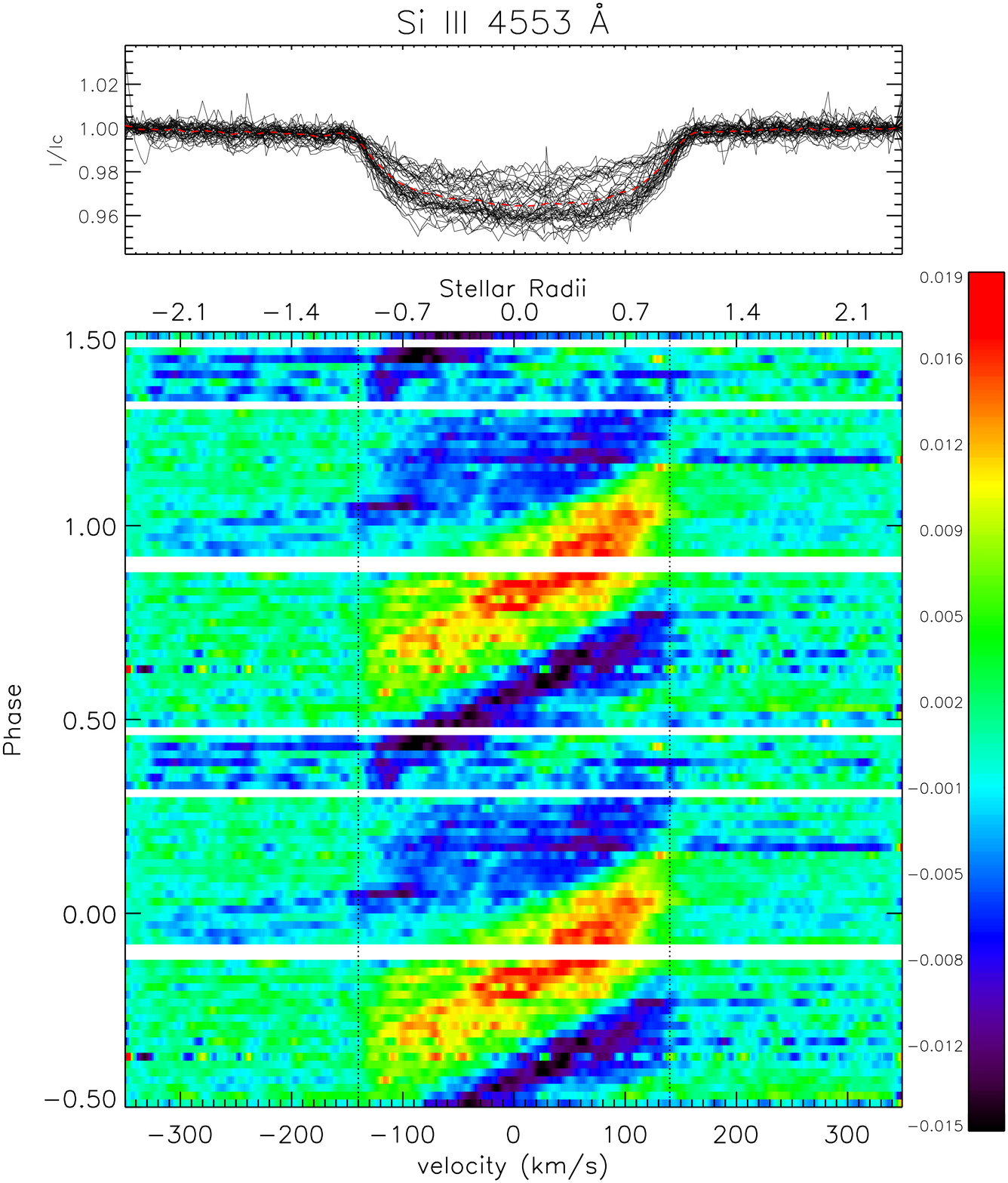}
\includegraphics[width=3.4 in]{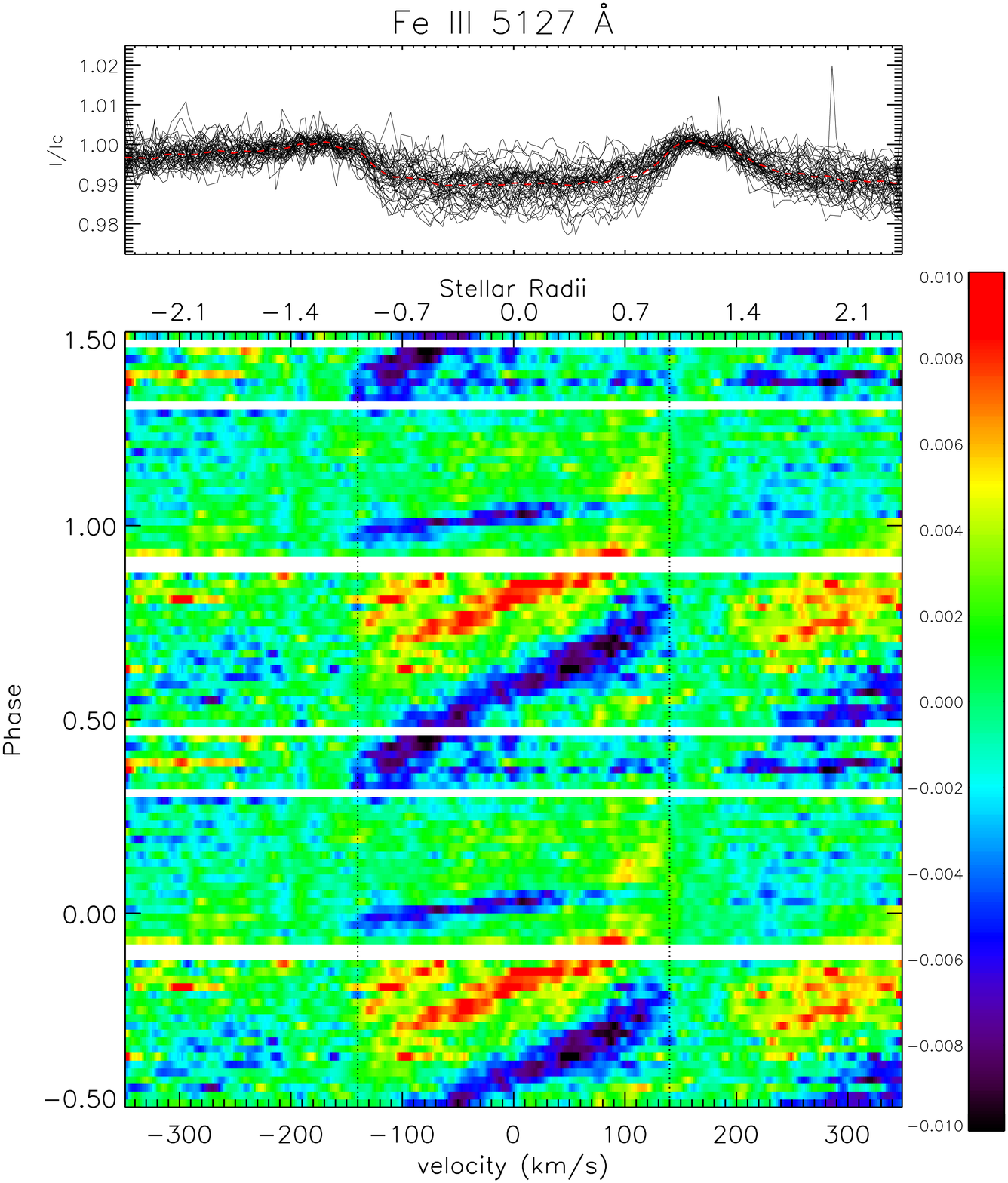}
\caption{Dynamic spectral plots for several spectral lines phased using the ephemeris of Townsend et al. (2010).   Spectra are plotted as a difference from the mean profile.  Dashed vertical lines mark the stellar rotational velocity.  FEROS data were not used for He~{\sc i} 6678 \AA~due to a bad pixel feature in the blue wing of the line.  \textbf{Top Left}: He~{\sc i} 6678 \AA.  \textbf{Top Right}: C~{\sc ii} 4267 \AA.  \textbf{Bottom Top}: Si~{\sc iii} 4553 \AA.  \textbf{Bottom Right}: Fe~{\sc iii} 5127 \AA.}
\label{dynplot}
\end{figure*}

The spectroscopic properties of $\sigma$ Ori E have been a popular research focus for decades, since the discovery of strong helium lines by Berger (1956).  With this new set of high-resolution spectra, our future goal is to model the photospheric spectral variations in different elements using Doppler imaging, as well as magnetospheric variability.  Here we examine the profiles and variability of the spectral lines of $\sigma$ Ori E in order to evaluate their qualitative properties, and identify lines appropriate for magnetic Doppler imaging (MDI; Piskunov \& Kochukhov 2002).  Due to the hot temperature of B2Vp stars, strong spectral lines in the optical spectrum are primarily H~{\sc i} and He {\sc i}.  We searched for weaker lines with high enough S/N and which showed variability.  Investigation of various lines of C~{\sc ii}, N~{\sc ii}, O~{\sc ii}, Al~{\sc iii}, Si~{\sc iii}, S~{\sc ii}, and Fe~{\sc iii} support the idea that the variability of $\sigma$ Ori E extends to both circumstellar and photospheric environments.  Each of the lines exhibits variability, although some on a more pronounced scale than others.   

In a study of optical line variability, Reiners et al. (2000) used the FEROS commissioning spectra of $\sigma$ Ori E (Kaufer et al. 1999) to study and model photospheric lines of He~{\sc i}, C~{\sc ii}, and Si~{\sc iii}.    The authors apply a model with a dipole magnetic field and two circular regions of variable abundance and size at the magnetic poles.  Reiners et al. (2000) conclude that for $\sigma$ Ori E the magnetic polar regions (each with a radius of 60\degr) are overabundant in helium ($\log$(He/H) = 0.0) and deficient in metals. Here, we present our own analysis of the same FEROS spectra, as well as the new spectropolarimetric data.  As this star varies on such a short time scale, we analyze the original intensity spectrum derived from each single sub-exposure, instead of the co-added spectra described in Sect. 2.  We limit our current study to H~{\sc i}, He~{\sc i}, C~{\sc ii}, Si~{\sc iii}, and Fe~{\sc iii}.

\subsection{Hydrogen}

In Figure 2 of Townsend et al. (2005), the H$\alpha$ FEROS spectra are presented in their comparison of RRM model predictions with observations.  Here, Fig. \ref{EW} plots the equivalent widths (EW) of the FEROS, ESPaDOnS, and Narval spectra for H$\alpha$, H$\beta$, and H$\gamma$ as a function of rotational phase.  Higher Balmer lines show trace amounts of circumstellar emission.  The dynamical plots of H$\alpha$ and H$\beta$ (Fig. \ref{hyddynplot}) shows the spectra minus a photospheric synthetic spectrum computed using a 23000~K, $\log g =4.0$ non-LTE model from TLUSTY (Lanz \& Hubeny 2007) together with SYNSPEC line formation code\footnote{http://nova.astro.umd.edu/Synspec43/synspec.html}, leaving only the assumed circumstellar profiles.  The subtraction of a single synthetic profile from these spectra, assumes that the hydrogen profile does not vary from any alternative processes (e.g. pulsation).  With the new spectra included, we confirm the double S-wave variation of circumstellar hydrogen.  The equivalent width curves for each of the Balmer hydrogen lines show maximum emission at phases 0.2 and 0.8, coincident with viewing the phases of magnetic extrema.  Maximum absorption occurs at phases 0.0 and 0.4, coincident with viewing the star's magnetic equator.  At these phases, we speculate that extra absorption may be due to occultation of the star by circumstellar material.  The secondary photometric minimum occurs at the same rotational phase, suggesting a small portion of the stellar disk is blocked by the co-rotating cloud of plasma.  The asymmetry of the red and blue shifted emission at maximum value confirms that the cloud emission measures are not equivalent.

\subsection{Helium}

It is well established that $\sigma$ Ori E is a He-strong star with patches of helium overabundance surrounded by regions of normal or deficient abundance (e.g. Veto 1991; Reiners et al. 2000).  Fig. \ref{EW} shows the equivalent width of the He~{\sc i} 5876~\AA~and 6678~\AA~lines as a function of rotational phase.  The dynamical plot in Fig. \ref{dynplot} displays the spectra as a difference from the mean profile.  Maximum absorption occurs at phase 0.4 and 0.9, whereas minimum absorption occurs at phases 0.2 and 0.6.  The maximum absorption at phase 0.9 is several times the absorption at phase 0.4.  The timing of the smaller absorption maximum corresponds with a phase of magnetic null, as well as a region where the H$\alpha$ emission is at a relative minimum.  The increase in absorption at phase 0.4 may then be due to occultation of the star by the circumstellar material, as suggested for hydrogen.   If this is the case, we would expect a similar effect at phase 0.0, however the effect should be much smaller than the observed absorption maximum at phase 0.9.

\subsection{Metals}

In addition to H and He, we also consider the C~{\sc ii} 4267 \AA, Si~{\sc iii} 4553 \AA, and Fe~{\sc iii} 5127 \AA~lines.  Fig. \ref{EW} displays the equivalent widths as a function of rotational phase for these lines.  Dynamical plots are shown in Fig. \ref{dynplot} with the spectra plotted as the difference from the mean profile.  All three lines vary similarly, showing an absorption minimum at phase 0.9.  This feature appears slightly wider in phase coverage in silicon, however with the error bars, any difference is difficult to detect.  Iron shows additional variability, although the spectral lines are weak and difficult to normalize.  In the dynamical plot of iron, a feature at phase 0.0 differs in movement from the rest of the features in these plots.  It is possible that this excess absorption is due to the same circumstellar effect as described for hydrogen and helium.   The minimum absorption of these metal lines at phase 0.9 coincides with maximum helium absorption.  This points to decreased metals in helium-strong regions, although it is unclear whether there is in fact a second absorption minimum to correspond to the helium maximum at phase 0.4.

\subsection{Summary}
New spectroscopic data confirm the diverse variability of optical spectral lines of $\sigma$ Ori E.  Hydrogen emission minima directly correspond to null longitudinal magnetic field measurements.  At these rotational phases (0.0 and 0.4), a cloud of plasma lies directly in front of the star.  The data indicate that these clouds are located at the intersection between the magnetic and rotational equators.  Regions of helium enhancement are coincident with regions of metal deficiency.  The longitudinal field maxima map to phases of maximum hydrogen emission.  At these phases, we view the star's magnetic poles with the plasma clouds in quadrature.  These new data are in good agreement with not only previous observations, but also the overall qualitative plasma geometry predicted by the RRM.

\section{Initial Confrontation}

\begin{figure}
\centering
\includegraphics[width=3.4 in]{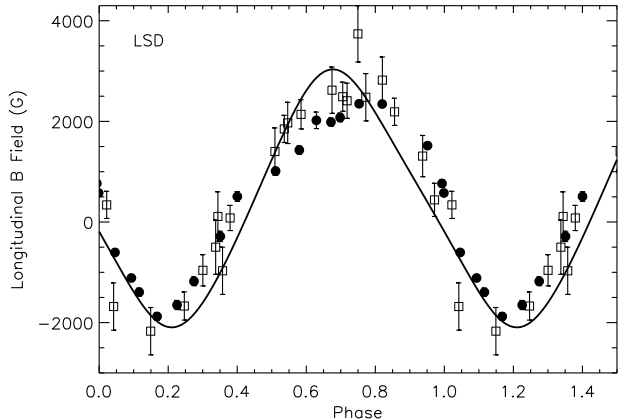}
\caption{Longitudinal magnetic field measurements (filled circles) for $\sigma$ Ori E, as well as data reported in Landstreet \& Borra (1978)  and Bohlender et al. (1987) (squares). Each data point is plotted with corresponding 1$\sigma$ error bars. The solid curve is the model longitudinal magnetic field curve from the offset dipole configuration of Townsend et al. (2005). The data are phased using the ephemeris of Townsend et al. (2010).}
\label{BlRRM}
\end{figure}

As a first exploitation of the current data set, we evaluate the appropriateness of the magnetic field configuration assumed by Townsend et al. (2005).  Prior to the present paper, the longitudinal magnetic field curve for $\sigma$ Ori E could be reasonably fit with a sinusoid, indicating a simple dipole structure.  Townsend et al. (2005) invoked a significant offset of the dipole magnetic field in their application of the RRM model to $\sigma$ Ori E in order to explain asymmetries in both the H$\alpha$ emission spectra and the photometric variations.  Using the ephemeris of Townsend et al. (2010), the offset dipole model fits the historical longitudinal field curve with a reduced $\chi^{2}$ of 2.92 (20 degrees of freedom; 22 data points - 2 model parameters).  The error bars on these data are large enough that they are not able to distinguish between a large range of magnetic field configurations.  When the new data are added (Fig. \ref{BlRRM}), the model reproduces the general character of the observations, but disagrees quantitatively with the detailed variation.  The offset dipole model fit to the new measurements computed from LSD profiles gives a reduced $\chi^{2}$ value of 92.9 (16 degrees of freedom).  The new H$\beta$ measurements fit with the same model give a reduced $\chi^{2}$ of 11.2 (16 degrees of freedom), while the new He line measurements have a reduced $\chi^{2}$ of 21.6 (34 degrees of freedom).  

\begin{figure*}
\centering
\includegraphics[width=5.8 in]{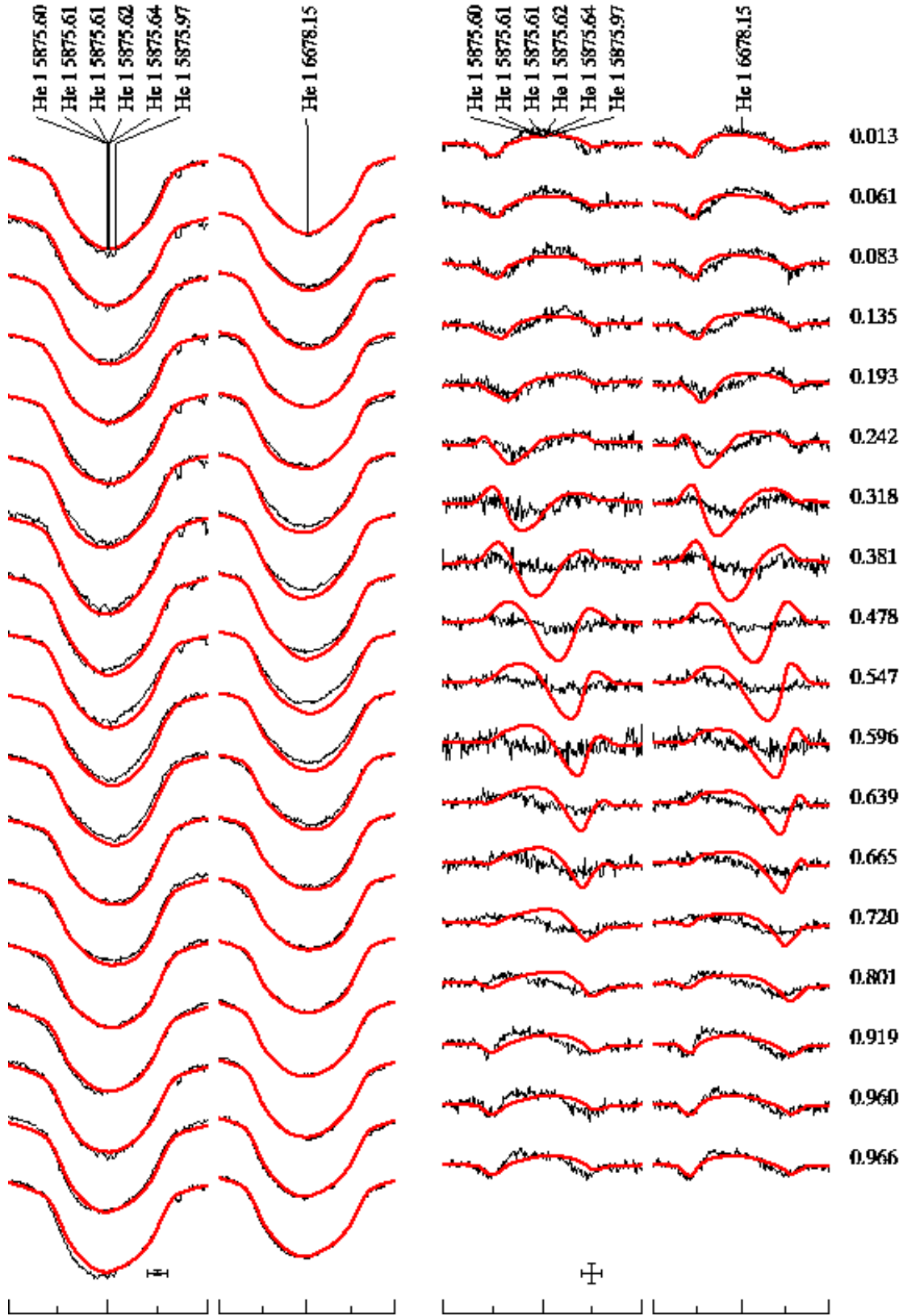}
\caption{Observed Stokes~$I$ and~$V$ profiles of the He {\sc i}~5876 and He~{\sc i} 6678 \AA~lines and synthetic spectra corresponding to the offset dipole configuration of Townsend et al. (2005).  There is a clear disagreement between model and data.  The phases noted are computed using the ephemeris HJD = 2442778\fd819 + 1\fd190842E.  The bars at the bottom of the plots give horizontal (1 \AA) and vertical (5\%) scales of the profile plots.}
\label{mdiprfdip}
\end{figure*}

Since the longitudinal magnetic field is an average quantity over the stellar disk, an even more stringent test of the RRM model magnetic field configuration can be obtained from the Stokes $I$ and Stokes $V$ profiles, which, by Doppler broadening, map variations of physical quantities across the surface of the star.  Fig. \ref{mdiprfdip} compares the predictions of the RRM offset dipole model with observed Stokes~$I$ and Stokes~$V$ profiles of the He~{\sc i} 5876 and 6678 \AA~lines. The synthetic Stokes parameters were calculated using the forward mode of the MDI code Invers10, adopting $v \sin i$=140 km~s$^{-1}$ (determined using the current spectra) and $i=75\degr$. In this specific model, the He abundance distribution was adopted from the Doppler imaging analysis of $\sigma$ Ori E, which will be presented in an upcoming paper. Nevertheless, models with uniform He distribution applied to the same magnetic field parameters produce similar profiles.  The magnetic field configuration of Townsend et al. (2005) predicts a far more extreme phase modulation and much larger amplitudes of the Stokes~$V$ profiles than are observed.  The disagreement between the synthetic and observed profiles is especially apparent for rotational phases 0.318-0.720.

Comparison of the RRM model magnetic field configuration and new spectropolarimetric data in both longitudinal magnetic field curve and Stokes~$I$ and Stokes~$V$ profiles demonstrate that the specific offset dipole field model invoked by Townsend et al. (2005) is not compatible with the new data.  The strong offset produces a much larger variation of the magnetic field across the visible stellar surface than is reflected by the observations, as illustrated by the model fit to the Stokes~$V$ profiles.  These profiles indicate that the model requires a very different field topology, further suggested by the periodic variability of the longitudinal magnetic field.  While we cannot definitely rule out the possibility of some alternate offset dipole, the asymmetric shape of the longitudinal magnetic field curve indicates a more complex field, requiring higher-order field components to match the shape of the variation.

\section{Summary and Future Work}

This paper lays the observational groundwork to re-evaluate the RRM model for $\sigma$ Ori E.  New spectroscopy is in agreement with previous observational studies of $\sigma$ Ori E (e.g. Pedersen \& Thomsen 1977; Landstreet \& Borra 1978; Reiners et al. 2000).  The H$\alpha$ variability confirms the general picture derived from the RRM model of a rapidly rotating star in which the magnetic field overpowers the stellar wind, allowing plasma to become trapped in a magnetosphere that co-rotates with the star.  While the RRM model agreed reasonably well with the previously observed longitudinal magnetic field curve for $\sigma$ Ori E, new, more precise measurements reveal a substantial variance between the shapes of the observed and modeled curves.  These results challenge the offset dipole form assumed in the Townsend et al. (2005) application of the RRM model to $\sigma$ Ori E, and indicate that future models of its magnetic field should also include complex, higher-order components.

As observations improve, more comprehensive modeling is required to explain the intricate physical phenomena of $\sigma$ Ori E.  We are currently in the process of using MDI to derive both abundance maps for strong variable spectral lines and magnetic field maps based on several He {\sc i} lines.  The circumstellar features observed in line profiles should not affect the resulting surface abundance maps, as these features have a much higher radial acceleration than the stellar surface (see e.g. Donati \& Collier Cameron 1997; Donati et al. 1999).  These new abundance maps will provide detailed information of the surface distribution of several elements, as well as provide the information to compute a total synthetic photospheric light curve for $\sigma$ Ori E.  We plan to use detailed model atmospheres to calculate flux distributions and simulate a light curve, as described by Krti\u{c}ka et al. (2007).  The magnetic field mapping will offer greater clarity into the structure of the magnetic field, however as our data do not include linear polarization (Stokes~$Q$ and~$U$ spectra) data, magnetic inversions require an assumption about global field topology to reduce ambiguity.  In its current state, the RRM model can only compute predictions for a dipole magnetic field.  We anticipate, with these future models, a clearer understanding of the remarkable physical phenomena of $\sigma$ Ori E.  Moreover, this work will provide the groundwork for analysis and modeling of similar massive magnetic stars.

\section*{Acknowledgments}

MEO acknowledges support from the NASA Delaware Space Grant, NASA Grant \#NNG05GO92H.  GAW acknowledges Discovery Grant support from the Natural Science and Engineering Research Council of Canada (NSERC), and from the Academic Research Program of the Royal Military College of Canada.  RHDT acknowledges support from NSF grants AST-0904607 and AST-0908688.  SPO acknowledges partial support from NASA ATP Grant \#NNX11AC40G.  OK is a Royal Swedish Academy of Sciences Research Fellow supported by grants from the Knut and Alice Wallenberg Foundation and the Swedish Research Council.

\label{lastpage}

\end{document}